\documentclass[showpacs,aps,prb,reprint,groupedaddress,twocolumn]{revtex4-1}

\usepackage{amsmath}
\usepackage{graphicx,subfigure}
\usepackage{psfrag}
\usepackage{color}
\usepackage[colorlinks]{hyperref}
\def\GFT{\overline{\bf G}}
\def\IT{\overline{\bf I}}

\def\fT{\overline{\bf f}}

\def\rrarg{\left(\mathbf{r},\mathbf{r}'\right)}

\def\ri{\left(\mathbf{r}_i\right)}

\def\br{\mathbf{r}}
\begin{document}



\title{Optical Forces Between Coupled Plasmonic Nano-particles near
Metal Surfaces and Negative Index Material Waveguides}


\author{C. \surname{Van Vlack}}\email[]{cvanvlack@physics.queensu.ca}
\author{P. Yao}
\altaffiliation[Current address: ]{Department of Optics and Optical Engineering, University of Science and Technology of China, 230026, P. R. China}
\author{S. Hughes}


\affiliation{Queen's University, Dept. of Physics, Kingston Ontario, Canada K7L 3N6}


\date{\today}

\begin{abstract}
We present a study of light-induced forces between two coupled plasmonic nano-particles above various slab geometries including a metallic half-space and a 280-nm thick negative index
material (NIM) slab waveguide. We investigate
optical forces by non-perturbatively calculating the scattered electric field via a Green function technique which includes the particle interactions to all orders.
For excitation frequencies near the surface plasmon polariton and slow-light waveguide modes of the metal and NIM, respectively, we find
rich light-induced forces and significant dynamical back-action effects. Optical quenching is
found to be important in both metal and NIM planar geometries,
which reduces the spatial  range of the achievable inter-particle forces.
However, reducing the loss in the NIM allows radiation to propagate through the slow-light modes more efficiently, thus causing the light-induced forces to be more pronounced between the two particles.
To highlight the underlying mechanisms by which the particles couple,
we connect our Green function calculations to various familiar quantities in quantum optics.
\end{abstract}



\maketitle

\section{Introduction \label{sec:intro}}

Since the proposal of using intense laser light to trap atoms~\cite{PhysRevLett.40.729} and particles~\cite{Ashkin-Science-210-1081} by Ashkin, scientists have achieved a remarkable ability to manipulate matter via optical forces~\cite{Grier-Nature-424-21}. This has lead to a plethora of light-matter force manipulation techniques, from Gaussian beam optical traps~\cite{Barton-JournalofAppliedPhysics-66-4594}, that are now routinely used in laboratories\cite{Neuman-RSI-75-2787}, to near-field nanometric tweezers~\cite{Novotny-PRL-79-645} which have motivated an entire field involving plasmonics to enhance and manipulate optical forces~\cite{Schuller-NatureMaterials-9-193,Quidant-LaserandPhotonicsReview-2-47,RevModPhys.82.1767}.
The use of plasmonic structures allow strong evanescent field enhancements near the plasmon resonances of the system, and can efficiently
trap particles with a size smaller than the wavelength of illumination. Examples include a patterned substrate with metallic particles~\cite{Righini-Nat.Phys.-3-477,Grigorenko-NATP-2-365W}, metallic near-field tips~\cite{Novotny-PRL-79-645}, or close to nano-apertures in thin metallic films~\cite{Juan-NaturePhysics-5-915}. In the latter case, Juan {\it et. al.}~\cite{Juan-NaturePhysics-5-915} recently examined the trapping of 50~nm and 100~nm polystyrene particles near a nano-aperture in a thin sheet of gold and illuminated with light very near to the transmission cutoff wavelength of the aperture. It was observed that as the particles were trapped in this geometry, the scattering induced by the particle worked to {\em enhance} the trapping, causing a self-induced ``back-action''.
This back-action illustrates that when considering the optical forces on nano-particles (NPs) near resonances, the NPs can interact {\em non-perturbatively} with the system and thus any models or theoretical  descriptions must include NP coupling in a self-consistent way.

\begin{figure}[b]
\vspace{-20pt}
\hspace{-25pt}\includegraphics[width=0.8\columnwidth]{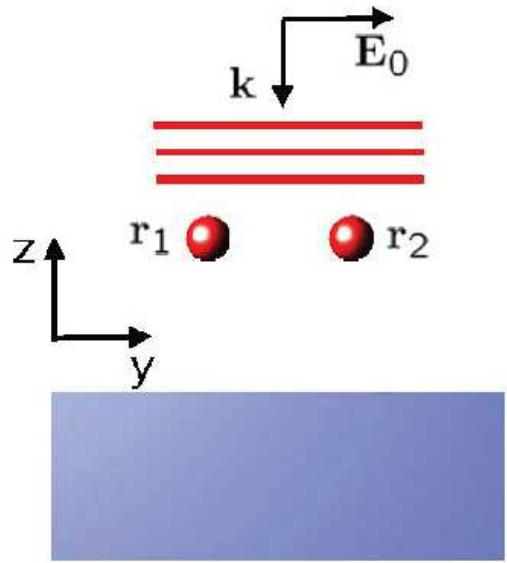}
\vspace{-10pt}
\caption{(Color online) Schematic of the geometry. We consider the optical forces on two NPs with various resonance frequencies and radii of $0.015 \lambda$ in air above a planar geometry.  We study both an infinite half-space of silver, and a 280~nm metamaterial slab which supports slow light modes (see text for details). The NPs are illuminated with a plane wave perpendicular to the interface, with the electric field polarization along the axis between the particles.  \label{fig:scheme}}
\end{figure}

Similar to field-enhancements using plasmonic structures, it is also possible to obtain
radiation enhancements using metamaterial structures which can be engineered with a
negative index of refraction ($\varepsilon<0$, $\mu <0$). Veselago introduced the concept of a negative refraction in 1968~\cite{Veselago-SPU-10-509}, and, in  2000, Smith {\it et. al.}~\cite{Smith-PRL-84-4184} experimentally realized such a material.
 This breakthrough initiated the field of ``metamaterials'' in which patterned materials with unit cells smaller than the wavelength of illumination were created to tune the effective material responses. Soon after, predictions were made of super-lensing~\cite{Pendry-PRL-85-3966}, cloaking devices~\cite{Pendry-SCI-312-1780,Schurig-SCI-314-977} and ``left-handed'' waveguides \cite{Shadrivov-PRE-67-057602}, all prompted by the ability to create negative index materials (NIMs).

Negative index materials also posses interesting quantum optics  and quantum electrodynamics (QED) properties.
 For instance, K\"{a}stel and Fleischhauer
\cite{Kastel-PRA-71-011804} examined an idealized (non-absorbing) negative index material placed on top of a mirror with an atom above the entire structure. They found that depending on the height of the atom it was possible to obtain complete suppression of spontaneous emission. Additionally, they examined atoms separated by a perfect negative index slab ($n=-1$) and showed that it was possible to obtain perfect subradiance and superradiance.
 Yao {\it et. al.}~\cite{Yao-PhysicalReviewB-80-195106} examined the enhancement of the spontaneous emission rate (Purcell Factor\cite{Purcell-PR-69-681}) above a NIM slab which supports a collection of slow-light modes in the region of negative index \cite{Shadrivov-PRE-67-057602, Lu2009}; related Purcell factor enhancements in NIM waveguides have been found
 by Xu {\em et al.}~\cite{Xu-PRA-79-043812} and Li {\em et al.} \cite{Li-PRB-80-045102}.
  In the slow light region, very large field enhancements can be obtained similar to the enhancements found in plasmonic structures. However, there were two important differences between the plasmonic and NIM systems: ($i$) the effects of quenching  through non-radiative energy transfer, and ($ii$) the role of the quasi-static approximation, i.e., the neglect of dynamic retardation effects. In plasmonic systems, it is possible to obtain strong Purcell factor enhancements however most of the emission is absorbed by the metal
  manifesting in non-radiative quenching~\cite{PhysRevLett.96.113002}. This optical quenching is a result of the intrinsic material loss which, in principle, can be tuned in metamaterial systems.
For metals, the quasi-static approximation is also typically used~\cite{PhysRevB.62.11185,Joulain-PRB-68-245405} for small objects approaching the surface ($kh<1$, where $k$ is the wavevector in the background medium containing the object and $h$ is the height above the surface),
 but for NIMs, such an approximation does not necessarily hold even for $kz\sim0.03$ \cite{Yao-PhysicalReviewB-80-195106}.

 Another potential advantage of NIMs, is that it is possible to obtain large Purcell factor enhancements in the optical region of spectrum, which also translates
to larger spatial distances.
 Recent experimental results~\cite{Noginov:10} have shown the possibilities of engineering the spontaneous emission rate using metamaterials composed of silver nano-wires embedded in PMMA. The nano-wire structure was shown to exhibit an in-plane anisotropic hyperbolic dispersion curve which have been shown to exhibit a negative refractive index\cite{PhysRevB.71.201101}. By exciting dye molecules embedded on top of the metamaterial the decay rate of the dye molecules was shown to be enhanced by a factor of 6 at $\lambda = 800$~nm compared to dye deposited on silver and gold films. This was even greater than expected when compared to a semi-classical model which predicted an enhancement of 1.8 \cite{2009arXiv0910.3981J}.

In this work, we theoretically investigate the optical forces on NPs above metallic and NIM planar geometries by self-consistently including the interaction of the NPs with the scattered electromagnetic field.
We exploit a photon Green function technique that can be used to introduce multiple particles within the dipole approximation~\cite{PhysRevB.64.035422} or, if required,  by using the full coupled dipole method~\cite{Purcell-AstrophysJ.-186-805,Draine-AstrophysJ.-333-848,Chaumet-Phys.Rev.B-61-14119--14127}, where
 the latter discretizes the NPs as small polarizable subunits. Green function and
 coupled dipole approaches have been very successful in examining light scattering~\cite{PhysRevB.58.2310} and optical forces~\cite{Chaumet-Phys.Rev.B-61-14119--14127,PhysRevB.64.035422} in dielectric particles located in evanescent fields above glass, and for computing optical forces between metallic particles in free space~\cite{Arias-Gonzalez:03} and above glass~\cite{PhysRevB.62.11185}. Green function approaches have also been used to successfully model optical trapping using near field optics~\cite{PhysRevLett.88.123601}.
In addition to computing the light-induced forces, we also analyze the properties of the Green function, both with and without the particles, which contains all the key electromagnetic interactions of the system, including coupling to surface plasmon polaritons (SPP) of the metallic half-space, the localized surface plasmons (LSP) of the particles, and evanescent coupling to slow light modes (SLMs) of the NIM slab. To help clarify the underlying physics, we also make direct  comparisons with various well known concepts
in quantum optics, such as the Purcell Factor, the Lamb shift, and real and virtual photon exchange; all
of these effects are relevant for understanding the ensuing coupling dynamics between the particles.

Our paper is organized as follow. In Sec.~\ref{sec:theory} we describe the theory
of light induced-scattering from NPs close to multi-layered surfaces, including, \ref{sec:GF_calc} -- the self-consistent calculation of the Green function and the electric field, and \ref{sec:forces} -- the calculation of the force from the total electric field. We exemplify our theoretical
results for silver half-space in
Sec.~\ref{sec:silver}, and for the NIM slab in Sec.~\ref{sec:metamaterial}. In Sec.~\ref{sec:conclusions}, we conclude.

\section{Theory \label{sec:theory}}

\subsection{Green Function Calculation \label{sec:GF_calc}}

For a spherical particle with radius $a$, and in the limit where $k_B a<<1$, the ``bare'' (i.e., no radiative coupling) polarizability is given by the Clausius-Mossotti relation,
\begin{equation}
\label{eq:alpha0}
  \alpha^0\left(\omega\right) = 4\pi \varepsilon_B a^3\frac{\varepsilon\left(\omega\right) - \varepsilon_B }{\varepsilon\left(\omega\right) + 2\varepsilon_B }\,,
\end{equation}
where $\varepsilon(\omega)$ is the particle dielectric constant (relative electric
permittivity) imbedded
in a homogeneous material with a background dielectric constant, $\varepsilon_B$,
assumed to be real. Here $k_B = \omega\sqrt{\varepsilon_B}/c$ is the wavevector in the background material.
It was shown by Draine~\cite{Draine-AstrophysJ.-333-848} that in order to satisfy the optical theorem, this polarizability must be corrected to include the {\em homogeneous-medium} contribution  to radiative reaction:
\begin{equation}
\label{eq:alpha}
  \alpha\left(\omega\right) = \frac{\alpha^0\left(\omega\right)}{1-\frac{3 \alpha_0\left(\omega\right) M\left(\omega\right)}{4 \pi \varepsilon_B a^3}}\,,
\end{equation}
where the self-induction term, $M(\omega)$, can be calculated exactly for a spherical particle~\cite{Martin-PRE-58-3909,Yaghjian-Proc.IEEE-68-248}, and for $k_B a<<1$ can be approximated as $M = 2i(a k_B)^3/9$.
Alternatively, this term comes naturally from the
homogeneous contribution of the Green function.

Our system is initially characterized by an initial electric field $\mathbf{E}^{(0)} \left(\mathbf{r};\omega \right)$, and an initial Green function $\GFT^{(0)} \left(\mathbf{r},\mathbf{r}';\omega \right)$, in the
{\em absence} of any particles. The field, $\mathbf{E}^{(0)} \left(\mathbf{r};\omega \right)$, can be of any form we wish (e.g., Gaussian beam, plane wave including reflections from surface), and we define it as the field prior to adding any scatterers. We subsequently introduce $N$ particles into the system where the $i$th particle is at position $\br_i$, with polarizability $\alpha_i\left(\omega\right)$. The total
electric field -- excitation field plus particle scattered field
 -- can be calculated self-consistently  from
\begin{equation}
\begin{split}
\label{eq:En}
  \mathbf{E}^{(N)}\left(\mathbf{r};\omega \right) &= \mathbf{E}^{(0)}\left(\mathbf{r};\omega \right) \\
&+ \sum_{i=1}^N \alpha_i\left(\omega\right) \GFT^{(0)} \left(\mathbf{r},\mathbf{r}_i;\omega \right) \cdot
\mathbf{E}^{(N)}\left(\mathbf{r}_i;\omega \right)\,.
\end{split}
\end{equation}
Similarly, the Green function of the system after adding $N$ particles can be calculated via the Dyson equation~\cite{Martin:94},
\begin{equation}
\label{eq:Gn}
\begin{split}
\GFT^{(N)}\left(\mathbf{r},\mathbf{r}'\omega \right) &= \GFT^{(0)} \left(\mathbf{r},\mathbf{r}';\omega \right) \\
&+ \sum_{i=1}^N \alpha_i(\omega) \GFT^{(0)}  \left(\mathbf{r},\mathbf{r}_i;\omega \right) \cdot \GFT^{(N)} \left(\mathbf{r}_i,\mathbf{r}';\omega \right)\, ,
\end{split}
\end{equation}
which is now the total Green function of the medium, including the
response of the NPs.
These equations form the basis of the coupled dipole method~\cite{Purcell-AstrophysJ.-186-805},
and, importantly, they apply to any general inhomogeneous and lossy media.

To help clarify the underlying physics we will further
 assume particles with a size much smaller than
 the wavelength, and  consider each NP within the dipole approximation; however it should be noted that at very short inter-particle distances this approximation eventually breaks down~\cite{doi:10.1021/jp911371r}.
Thus we will restrict the distances to regimes
where the dipole approximation is expected
to be a good approximation;  in this way, we include the particle dipoles
 exactly, while essentially dealing with
spatially-averaged particle quantities.
Using the Dyson equation, we can also rewrite the right hand side of Eq.~(\ref{eq:En}) to be given
only in terms of the excitation field~\cite{Martin-PRE-58-3909}, $\mathbf{E}^{(0)}\left(\mathbf{r};\omega \right)$. One has
\begin{equation}
\begin{split}
\label{eq:En2}
  \mathbf{E}^{(N)}\left(\mathbf{r};\omega \right) &= \mathbf{E}^{(0)}\left(\mathbf{r};\omega \right) \\
&+ \sum_{i=1}^N \alpha_i\left(\omega\right) \GFT^{(N)} \left(\mathbf{r},\mathbf{r}_i;\omega \right) \cdot
\mathbf{E}^{(0)}\left(\mathbf{r}_i;\omega \right)\, ,
\end{split}
\end{equation}
where $\GFT^{(N)}$  includes the particle(s) response.
The Green function, $\GFT$, can also be separated into homogeneous (direct), $\GFT_{\rm hom}$, or scattered (indirect), $\GFT_{\rm scatt}$, contributions.
Since ${\rm Re}[\GFT_{\rm hom}({\bf r},{\bf r}'\rightarrow{\bf r})]$ diverges, in what follows below
we will consider the non-divergent  ${\rm Re}[\GFT_{\rm scatt}({\bf r},{\bf r}')]$  when
${\bf r}={\bf r}'$, as this is the only relevant {\em photonic} contribution.
While ${\rm Re}[\GFT_{\rm hom}({\bf r},{\bf r})]$ can give a very small
{\em vacuum} Lamb shift, this effect can be already included
by simply redefining the resonance frequency of the particles.
In addition, since $\alpha(\omega)$ includes
the effect of ${\rm Im}[\GFT_{\rm hom}({\bf r},{\bf r})]$
through the self-induction term, then we need
only consider $\GFT_{\rm scatt}({\bf r},{\bf r})$, i.e.,
when ${\bf r}={\bf r}'$. For ${\bf r} \neq {\bf r}'$,
we consider the full $\GFT({\bf r},{\bf r})= \GFT_{\rm scatt}({\bf r},{\bf r}')
+ \GFT_{\rm hom}({\bf r},{\bf r}')$.
%
%
Frequently, it is possible to solve Eqs.~(\ref{eq:En}-\ref{eq:En2}),  {\em perturbatively}, by considering all $\GFT^{(N)}$ and ${\bf E}^{(N)}$ on the RHS to be the {\em unperturbed} quantities, i.e.,
approximated by $\GFT^{(0)}$ and ${\bf E}^{(0)}$. When the system constituents are far separated this can hold; however, as the particles come closer together and nearer the planar surface,
we will show that dynamical coupling  become important and cannot be neglected.

To examine the mechanisms of light-induced forces,
 it is useful to consider the Green function of the planar surface with and without the inclusion of the particle(s). We define the  complex \emph{local density of states} (LDOS) as
\begin{equation}
\label{eq:rhor}
  \rho^{(N)}_{m} \left(\mathbf{r};\omega \right)= \frac{ G^{(N)}_{mm}\left(\mathbf{r},\mathbf{r};\omega \right)}{{\rm Im} \left[G^{\rm hom}_{mm}\left(\mathbf{r},\mathbf{r};\omega \right)\right]}\,,
\end{equation}
where $G_{mm}^{(N)}\left(\mathbf{r},\mathbf{r};\omega\right)$ is given by Eq.~(\ref{eq:Gn}).
Although the LDOS only depends on
$ {\rm Im}[\rho^{(N)}_{m}]$, we introduce
$\rho^{(N)}_{m}$ as a complex quantity for ease of notation.
For example,
the imaginary part of the Green function in a homogeneous lossless material at $\br = \br'$ is given by
\begin{equation}
\label{eq:Ghom}
\begin{split}
{\rm Im} \left[G^{\rm hom}_{mm}\left(\mathbf{r},\mathbf{r};\omega\right)\right] = \frac{\omega^3 \sqrt{\varepsilon_B}}{6\pi c^3} ,
\end{split}
\end{equation}
which is  related to the homogeneous-medium LDOS.
When the homogeneous material contains loss, then both the real and imaginary part of $\GFT_{\rm hom}$ formally diverge as $\br \rightarrow \br'$ instead of just the real part of $\GFT_{\rm hom}$ (i.e.,
as in the case of a lossless material).
Consistent with our discussions and notation above,
when $\rho^{(N)}_{m} \left(\mathbf{r};\omega \right)=0$, this means that the LDOS  at that point is equal to the homogeneous density of states, as the scattered contribution is zero;
thus the total $\rho^{\rm tot}_{m} = \rho^{}_{m} +1$,
and we will focus on $\rho^{}_{m}$.
Since we will consider the force interaction between
two particles, it is also useful
to introduce a complex \emph{non-local density of states} (NLDOS),
\begin{equation}
\label{eq:rhorr}
  \rho^{(N)}_{mn} \left(\mathbf{r},\mathbf{r}';\omega \right)= \frac{G_{mn}^{(N)}\left(\mathbf{r},\mathbf{r}';\omega \right)}{{\rm Im} \left[G_{mm}^{\rm hom}\left(\mathbf{r}',\mathbf{r}';\omega \right)\right]},
\end{equation} \\
which describes light propagation between the
two space points ${\bf r}$ and ${\bf r'}\neq {\bf r}$.

Many of these quantities are useful also for connecting to
the quantum optical properties of the optical NPs~\cite{Novotny-Nano-Optics,Vogel-Quantum-Optics}.
This is a key strength of the Green function approach over brute-force numerical electromagnetic techniques
such as, e.g.,  FDTD (finite-difference time domain) \cite{Taflove-FDTD}.
In the case of the complex LDOS, the real part of Eq.~(\ref{eq:rhor}) can describes frequency shifts (Lamb shifts) of an emitter caused by the environment, whereas the imaginary part describes the
 material-dependent spontaneous emission.
For the {\em photonic} Lamb shift,  one has 
 \begin{equation}
\label{lamb}
\delta\omega^{(N)} ({\bf r},\omega) = -\frac{{\bf d} \cdot {\rm Re}[\GFT_{\rm scatt}^{(N)}({\bf r},{\bf r};\omega)] \cdot {\bf d}}{\hbar \varepsilon_0}  ,
\end{equation}
for an emitter at position ${\bf r}$.
For calculations of light-induced optical forces, the real part of the LDOS can therefore manifest itself as resonance shifts of the local electric field.

The variation of the imaginary part of the LDOS causes gradient forces on the particle as it moves through the field.
In terms of non-local QED interactions  between two particles,
 the real part of the NLDOS describes virtual (instantaneous) photon exchange between two points or emitters and the imaginary part describes real (dynamic) photon exchange. Virtual photon exchange manifests itself in well known processes like F\"{o}rster coupling, which, for a homogeneous medium, exhibit an $R^{-3}$ scaling with inter-particle separation ($R$) in free space, and real photon exchange manifests itself in dipole-dipole coupling which exhibits an $R^{-1}$ dependence~\cite{Thomas-Phys.StatusSolidiB-230-25}. The effect of the real part of the NLDOS on classical optical forces manifests itself in the scattered contribution to the force and the imaginary part would contribute in a manner similar to radiation pressure. For a non-homogeneous medium,
 we stress that the photon coupling mechanisms
 are considerably more complicated than simple  F\"{o}rster coupling. In addition,
 we fully include {\em dynamical} retardation effects through the frequency-dependence
 of the response functions.

 The above prescriptions are relatively straightforward provided
 one knows the bare Green functions without any particles.
 For the calculation of the initial planar Green function, we use a
  well established multilayer technique~\cite{Sipe-JOSAB-4-481},  which is outlined in Appendix~\ref{sec:slab_GF} and further
  numerical details are detailed by Paulus {\it et. al.}~\cite{Paulus-PRE-62-5797}.
  Although we specialize our study for two NPs, the generalization to
  any arbitrary number of NPs is straightforward with no change
  in theoretical formalism.

\subsection{Light-Induced Forces \label{sec:forces}}

Using the electric field $\mathbf{E}^{(N)}\left(\mathbf{r}_i;\omega \right)$
at the NP position ${\bf r}_i$, the time-averaged total force on a particular NP from a time-harmonic electromagnetic wave
is~\cite{Chaumet-OLETT-25-1065} ($\omega$ is implicit),

\begin{widetext}
\begin{equation}
\begin{split}
 \left<\mathbf{F}\ri \right> = \frac{\varepsilon_0}{4T} \int_{-T/2}^{T/2}\left[\left(\alpha_i \mathbf{E}^{(N)}\ri\right.\right. +&\left.\left.\alpha^*_i \left(\mathbf{E}^{(N)}\ri\right)^* \right)\cdot \nabla
\left(\mathbf{E}^{(N)}\ri +\left(\mathbf{E}^{(N)}\ri\right)^* \right) \right.\\
 +& \left.\left(\alpha_i \dot{\mathbf{E}}^{(N)}\ri +\alpha_i^* \left(\dot{\mathbf{E}}^{(N)}\ri\right)^* \right) \times\left(\mathbf{B}^{(N)}\ri +\left(\mathbf{B}^{(N)}\ri\right)^* \right)\right],
 \end{split}
\end{equation}
\end{widetext}
where $\varepsilon_0$ is the permittivity of free space, $\mathbf{B}$ is the magnetic field and $T$ is the period of the time-harmonic radiation.
Upon carrying out the integration, and using $\mathbf{B} = 1/i\omega \, \nabla \times \mathbf{E}$, and $\dot{\mathbf{E}} = - i \omega \mathbf{E}$,
\begin{equation}
\begin{split}
 \left<F_j\right> = &\frac{\varepsilon_0}{2} {\rm Re} \left\{\alpha \left[ E^{(N)}_k \partial_k \left(E^{(N)}_j\right)^*  \right.\right.\\
+& \left.\left. \varepsilon^{jkl} \, \varepsilon_{lmn} E^{(N)}_k  \partial_m \left(E^{(N)}_n\right)^* \right]\right\} ,
 \end{split}
\end{equation}
where $\varepsilon^{jkl}$ is the  Levi-Civita tensor.
Using the relation $\varepsilon^{jkl} \, \varepsilon_{lmn} = \delta_{jm} \delta_{kn} - \delta_{jn}\delta_{km}$,
 we obtain  the desired force
\begin{equation}
\label{eq:F}
  \mathbf{F}\left(\mathbf{r}_i\right) = \frac{\varepsilon_0}{2} \sum_{j={x,y,z}} {\rm Re}\left[\alpha_i E^{(N)}_j\left(\mathbf{r}_i\right) \nabla \left(E^{(N)}_j\left(\mathbf{r}_i\right)\right)^*\right].
\end{equation}
This light-induced force describes the force on the particle due to the electric field and its interaction with the planar geometry as well as the scattering due to the other NPs in the system. This is different from the usual {\em gradient force} given by
$  \mathbf{F} = \frac{1}{2} \alpha_0 \nabla \left|\mathbf{E}\right|^2,$
which is only applicable when $\alpha$ is real and the phase of the field varies slowly in space.
\begin{figure}[t!]
\includegraphics[width=3.25in]{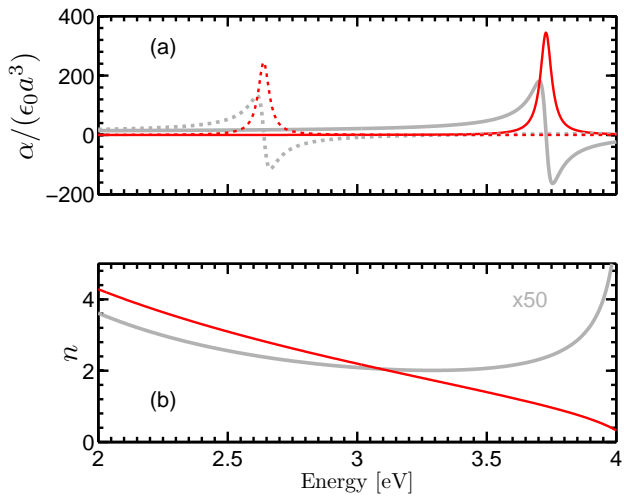}
\caption{(Color online) Particle and silver half-space response functions.
For both the silver half-space and the silver NP permittivity we use the Drude model [Eq.~(\ref{eq:drude})], with $\gamma = 51$ meV and $\varepsilon_r = 6$. For the silver half-space, the plasma frequency is given by $\omega_{pe} = 9.87$ eV. To tune the LSP of the NP to the SPP we set the plasma frequency at $\omega_{pe} = 10.56$ eV; and to tune the LSP of the NP to be off-resonant with the SPP, we set the plasma frequency at $\omega_{pe} = 7.47$ eV.
(a) Bare polarizability [Eq.~(\ref{eq:alpha0})] of a silver NP with radius $a = 5$ nm ($= 0.015 \, \lambda_{SPP}$) with the LSP resonance tuned to the SPP of the silver half-space, $\omega_{SPP} = \omega_{LSP} = 3.73$ eV (solid lines), and with the LSP resonance tuned to be off resonant with the SPP of the silver half-space, $\omega_{LSP} = 2.63$ eV (dashed lines). Gray-light curve indicates real parts and red-dark curve indicates imaginary parts. (b) Refractive index of the silver half-space, gray-light curve indicates real part
(scaled by a factor of 50) and red-dark curve indicates imaginary part.
\label{fig:index_silver}}
\end{figure}

\section{Silver Half-Space \label{sec:silver}}

 To calculate the optical forces on the NPs, we consider the geometry shown in Fig.~\ref{fig:scheme}, with two NPs in air above a planar structure. For silver, we consider a half-space geometry or an optically thick slab with a permittivity given via the Drude model,
\begin{equation}
\label{eq:drude}
 \varepsilon \left(\omega\right) = \varepsilon_r - \frac{\omega_{pe}^2}{\omega^2 + i\omega\gamma}.
\end{equation}
Here $\varepsilon_r = 6$ is the permittivity as $\omega \rightarrow \infty$, $\omega_{pe} = 9.87$ eV is the electric plasma frequency, and $\gamma = 51$ meV is the damping rate due to collisions\cite{Liu:08,Johnson-PRB-6-4370}. For a metallic half-space, the characteristic surface plasmon polariton (SPP) frequency is given by ${\rm Re}\left[\varepsilon\left(\omega=\omega_{SPP}\right)\right] = - \varepsilon_B$. Below this frequency, SPPs are confined to the interface and can only be coupled to by breaking the symmetry of the system (e.g., via a grating coupler). For a silver/air interface the SPP is located at $\omega_{SPP} = 3.73$ eV ($\lambda_{SPP} = 332$\,nm). Surface plasmon polaritons are well known for their ability to enhance the electric field in their vicinity, but these enhancements are also associated with high losses meaning
that coupling between objects or the far field can be suppressed/quenched.

For a spherical NP, the localized surface plasmon resonance is at ${\rm Re}\left[\varepsilon\left(\omega\right)\right] = -2 \varepsilon_B$  [see Eq.~(\ref{eq:alpha0})] which occurs at $\omega_{\rm LSP} = 3.49$~eV ($\lambda_{LSP} = 355$~nm) using the parameters for silver and air. To investigate coupling between the NPs and the resonances of the metallic half-space, we will fix $\varepsilon_r$ and $\gamma$ to  the parameters for bulk silver despite deviations from bulk-like behavior for small NPs \cite{RevModPhys.58.533}.
 For small NPs, the localized surface plasmon of the NP becomes strongly dependent on
 size~\cite{mcmahon:097403} and shape~\cite{Ye-LANG-25-1822} and can be further detuned by adding dielectric or metallic coatings~\cite{Zhang-Nano.Lett.-9-4061}. We shall use this as motivation to tune the LSP of our NPs
 to be either on resonance with the SPP ($\omega_{LSP} = \omega_{SPP} = 3.73$ eV, $\omega_{pe} = 10.56$ eV), or off resonance with the SPP ($\omega_{LSP} = 2.63$ eV, $\omega_{pe} = 7.47$eV). The bare polarizability and metal
  half-space permittivity response functions are shown in Fig.~\ref{fig:index_silver}.

In the following, we consider the geometry shown in Fig.~\ref{fig:scheme}, where NPs with tunable LSPs are located above a planar structure. We first use particles with a  radius of $0.015 \, \lambda_{SPP}$ ($=5~$nm) so as to have a reasonable sized particle
for which the dipole approximation will apply. We  vary the height, $h$, of the particles above the half-space (measured from the center of the particles) while keeping both particles at the same height for simplicity. Also, we vary the separation, $s$, between the particles (along the $y$ direction and measured from their centers) and the frequency of light illumination. To illuminate the particles, we use a homogeneous excitation field, which is a solution to the scattering problem
(incident light plus scattered light) without any NPs;  we choose the polarization to be along the direction of the particles ($y$ direction) to maximize the effect of the coupling between the particles.
To excite surface plasmons, which only exist for TM polarization, the symmetry of the system must be broken to allow coupling between the surface plasmon and the particles, so only the scattered field from the NPs can excite the SPP.
The incident intensity is 1 W/$\mu$m$^2$, which we choose only as a convenient reference --
the force scales linearly with the incident intensity as can be seen from Eq.~(\ref{eq:F}).
For comparison,
we note that the earth's gravitational force
on a 5~nm silver particle is 54~fN; for the intensities considered below, the gravitational force is smaller by a factor of $10^{-3}$, and thus can be safely neglected.

\begin{figure}[t!]
\centering
\subfigure[\,LDOS]{
\includegraphics[scale=0.4]{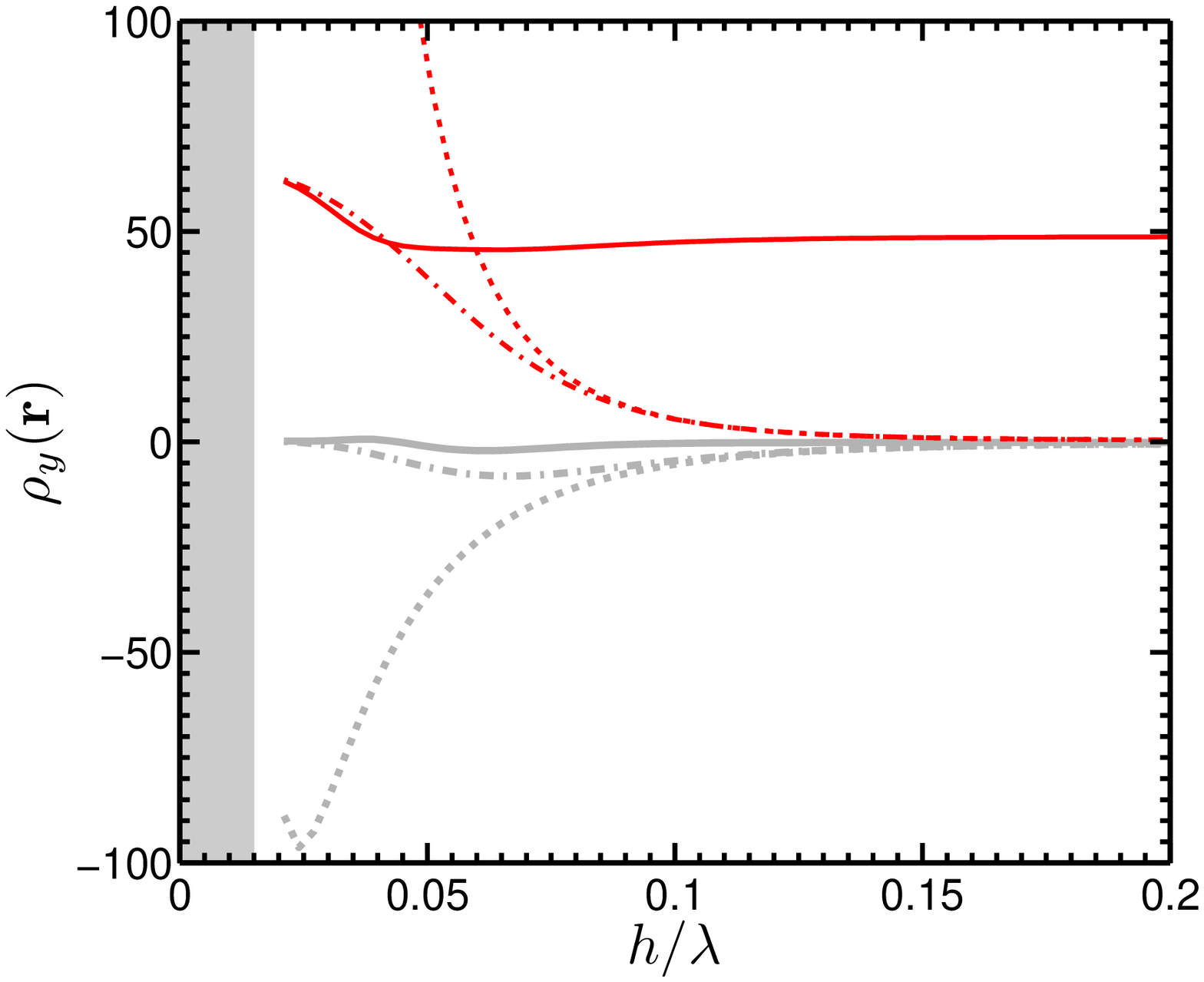}
}
\hfill
\subfigure[\,NLDOS ]{
\includegraphics[scale=0.4]{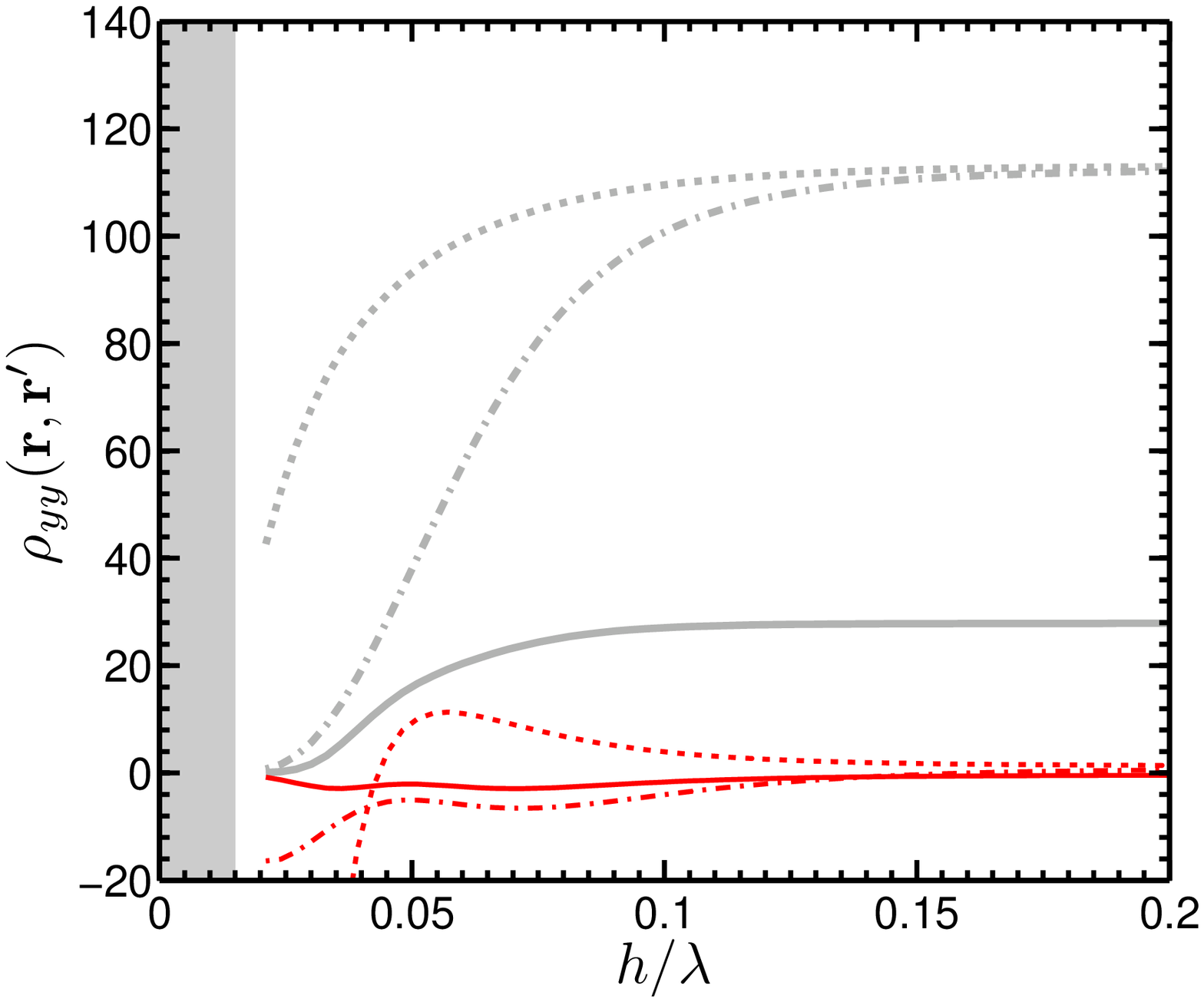}
}
\vspace{-0.2cm}
\caption{ (Color online)
(a) Complex LDOS, $\rho^{(N)}_{y}\left(\mathbf{r}', \omega\right)$, and (b) NLDOS,
$\rho^{(N)}_{yy}\left(\mathbf{r},\mathbf{r}', \omega\right)$, for a silver half-space as a function of height at $\omega = \omega_{SPP}$. For the LDOS, $\mathbf{r} = \left(0,0,h\right)$, and for the NLDOS, $\mathbf{r}' = \left(0,0,h\right)$, $\mathbf{r} = \left(0,0.05\,\lambda,h\right)$. Gray-light curves represent the real part and red-dark curves represent the imaginary part. The dashed line has no particles [$\rho^{(0)}_{y}\left(\mathbf{r};\omega\right)$, $\rho^{(0)}_{yy}\left(\mathbf{r},\mathbf{r}';\omega\right)$],
the chain line has one particle located at the position the LDOS/NLDOS is calculated [$\rho^{(1)}_{y}\left(\mathbf{r};\omega\right)$, $\rho^{(1)}_{yy}\left(\mathbf{r},\mathbf{r}';\omega\right)$ and $\mathbf{r}_1 = \left(0,0,h\right)$], and the solid line has two particles -- one where the LDOS/NLDOS is calculated aand the other at the same height as the first but separated by $0.05 \, \lambda$ in the $y$ direction [$\rho^{(2)}_{y}\left(\mathbf{r};\omega\right)$, $\rho^{(2)}_{yy}\left(\mathbf{r},\mathbf{r}';\omega\right)$, $\mathbf{r}_1 = \left(0,0,h\right)$, and $\mathbf{r}_2 = \left(0,0.05\,\lambda,h\right)$]. The gray shaded region indicates region where particles would overlap the planar structure.
 \label{fig:GF_silver}}
\end{figure}

In Fig.~\ref{fig:GF_silver} (a) we examine the $y$ component of the LDOS of the half-space, which will dominate the forces for the particular illumination scheme that we have selected. We vary the height [$\mathbf{r} = \left(0,0,h\right)$] keeping in mind that we cannot approach closer to the structure than our particle radius which is indicated by the gray shaded region. We consider $\omega = \omega_{SPP} = \omega_{LSP}$ and add the first particle at $\mathbf{r}_1 = \mathbf{r}$ and the second particle at $\mathbf{r}_2 = \left(0,0.05\, \lambda_{SSP},h\right)$. When there are no particles in the system, we can see that the imaginary part of the LDOS (red-dark dashed curve) diverges, which would lead to an infinite LDOS for an excited emitter; although this
 effect may seem surprising, such a divergence always happens above a lossy structure~\cite{Yao-PhysicalReviewB-80-195106,GayBalmaz200037} [Eq.~(\ref{eq:mult_G_tot_slab_QS})].
 However, the inclusion of the particle where we calculate the LDOS (red-dark chain curve) acts to renormalize the LDOS for distances $< 0.08 \,\lambda$. The addition of a second particle at the same height as the first, but separated
 by $0.05\, \lambda$ (center to center), further renormalizes the LDOS (red-dark solid curve), though
 it becomes apparent that the effect of the silver half-space becomes negligible for heights greater
 than about $0.06\,\lambda$ when there are two particles.

For the real part of the $y$ component of the LDOS (gray-light curves), and with no particle in the system (dashed), we see that there is a minimum as the particle approaches the half-space which would give a blue shift for an emitter placed close to the surface. Including a particle at this location (gray-light chain curve) reduces the blue shift and including the second particle (gray solid curve) causes a change in sign which means an emitter would be shifted to the red. For both the real and imaginary part of the LDOS these shifts are only seen by going beyond the perturbative limit and solving Eq.~(\ref{eq:Gn}) exactly. These results emphasize the pronounced
back-action effects that occur in describing the electromagnetic properties of
the medium.

\begin{figure}[h!]
\includegraphics[width=8.4cm] {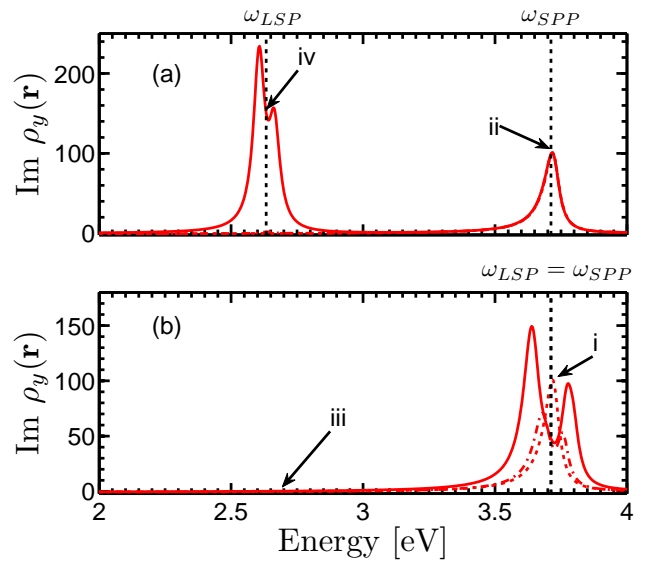}
\caption{(Color online) Im[$\rho^{(N)}_{y}\left(\mathbf{r}, \omega\right)$] for a silver half-space as a function of frequency at the position $\mathbf{r} = \left(0,0,0.05\,\lambda_{SPP}\right)$.
 The dashed line has no particles, Im[$\rho^{(0)}_{y}\left(\mathbf{r};\omega\right)$],
 the chain line has one particle, Im[$\rho^{(1)}_{y}\left(\mathbf{r};\omega\right)$], and
 the solid line has two particles, Im[$\rho^{(2)}_{y}\left(\mathbf{r};\omega\right)$] at locations $\mathbf{r}_1 = \left(0,0,0.05\lambda_{SPP}\right)$, and $\mathbf{r}_2 = \left(0,0.05\lambda_{SPP},0.05\lambda_{SPP} \right)$. (a) The particle LSPs are red detuned by $\Delta \omega = 1.1$~eV compared to the SPP frequency (see Fig. \ref{fig:index_silver}). (b) The particle LSPs are at the SPP frequency. Arrows indicate the particular scenarios we are examining in force graphs: `i' -- Fig.~\ref{fig:Force1}(a), `ii' --
 Fig.~\ref{fig:Force1}(b), `iii' -- Fig.~\ref{fig:Force1}(c) and `iv' -- Fig.~\ref{fig:Force1}(d).
 \label{fig:GF_freq1}
 }
\end{figure}

Figure \ref{fig:GF_silver} (b) shows the $yy$ component of the NLDOS in a similar manner to the LDOS described above, with $\omega = \omega_{SPP} = \omega_{LSP}$, $\mathbf{r}' = \mathbf{r}_1 =\left(0,0,h\right)$, and we consider $\mathbf{r} =\mathbf{r}_2 = \left(0,0.05\,\lambda_{SPP},h\right)$. With no particles in the system, the imaginary part of the NLDOS  (red-dark dashed curve) becomes large but remains finite as the surface is approached, implying that photons are very easily transferred from $\mathbf{r}'$ to $\mathbf{r}$ ({\em off scale}). However once a particle is added (red-dark chain curve), the NLDOS reduces drastically as the particle breaks the symmetry of the system allowing quenching to occur. Interestingly, the addition of the second particle (red-dark solid curve) further breaks the symmetry of the system and allows light to couple to more channels in the planar structure, which further enhances the quenching effect and reduces the NLDOS. This dramatic reduction is a result of the non-perturbative coupling between the particles and the half-space which is
 theoretically described through the self-consistent solution of Eq.~(\ref{eq:Gn}). For heights greater than $0.15\, \lambda$, light propagates purely via virtual photon propagation as the real part (gray-light curve) approaches a finite value (the homogeneous Green function) for both zero (dashed), and one (chain) particle. For two particles this happens even closer to the surface ($\approx 0.1 \, \lambda$) due to the additional scattering events. Real photon propagation occurs when the half-space begins to interact with the system and multiple paths are possible for a photon to reach $\mathbf{r}$ from $\mathbf{r}'$.

The {\em quasi-static} approximation is often invoked for particles very close to a surface or to each
other~\cite{GayBalmaz200037}, and this approximation holds for the imaginary part of the LDOS as the surface is approached at the SPP frequency; the real part deviates significantly in this limit. However, when the incident frequency is detuned  from the SPP  ($\omega = 2.63$~eV),  the quasi-static approximation again becomes valid. Similar results are found for the NLDOS, except when the inter-particle separation is greater than $0.07\,\lambda$ and the quasi-static approximation again breaks down. This means that for resonance interactions one must be very careful
about applying a quasi-static approximation.

\begin{figure*}[ht]
\centering
\subfigure[\,Force lines on particle 2, at region-`i' on Fig.~\ref{fig:GF_freq1}]{
\includegraphics[scale=0.4]{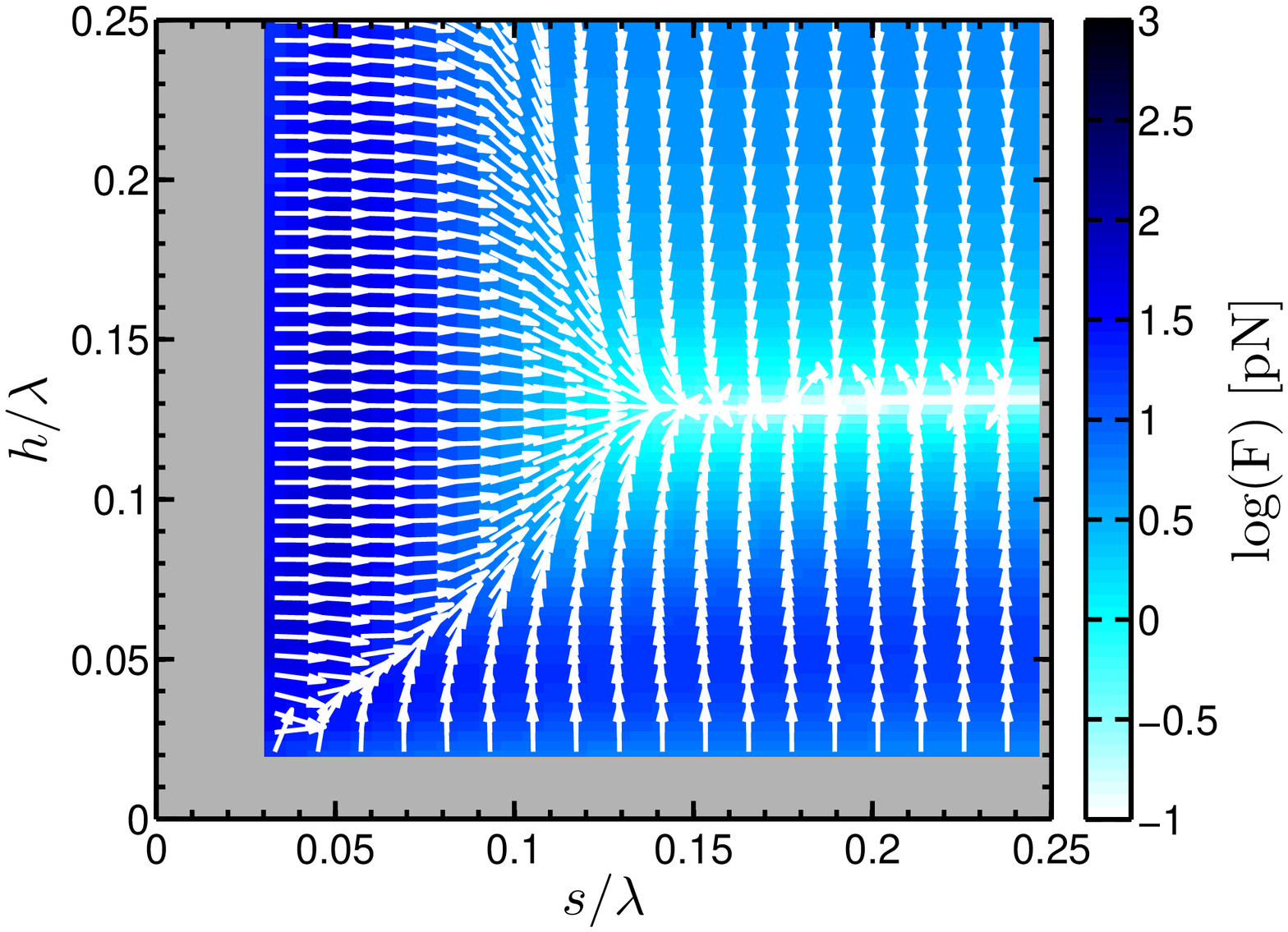}
}
\hfill
\subfigure[\,Force lines on particle 2, at region-`ii' on Fig.~\ref{fig:GF_freq1}]{
\includegraphics[scale=0.4]{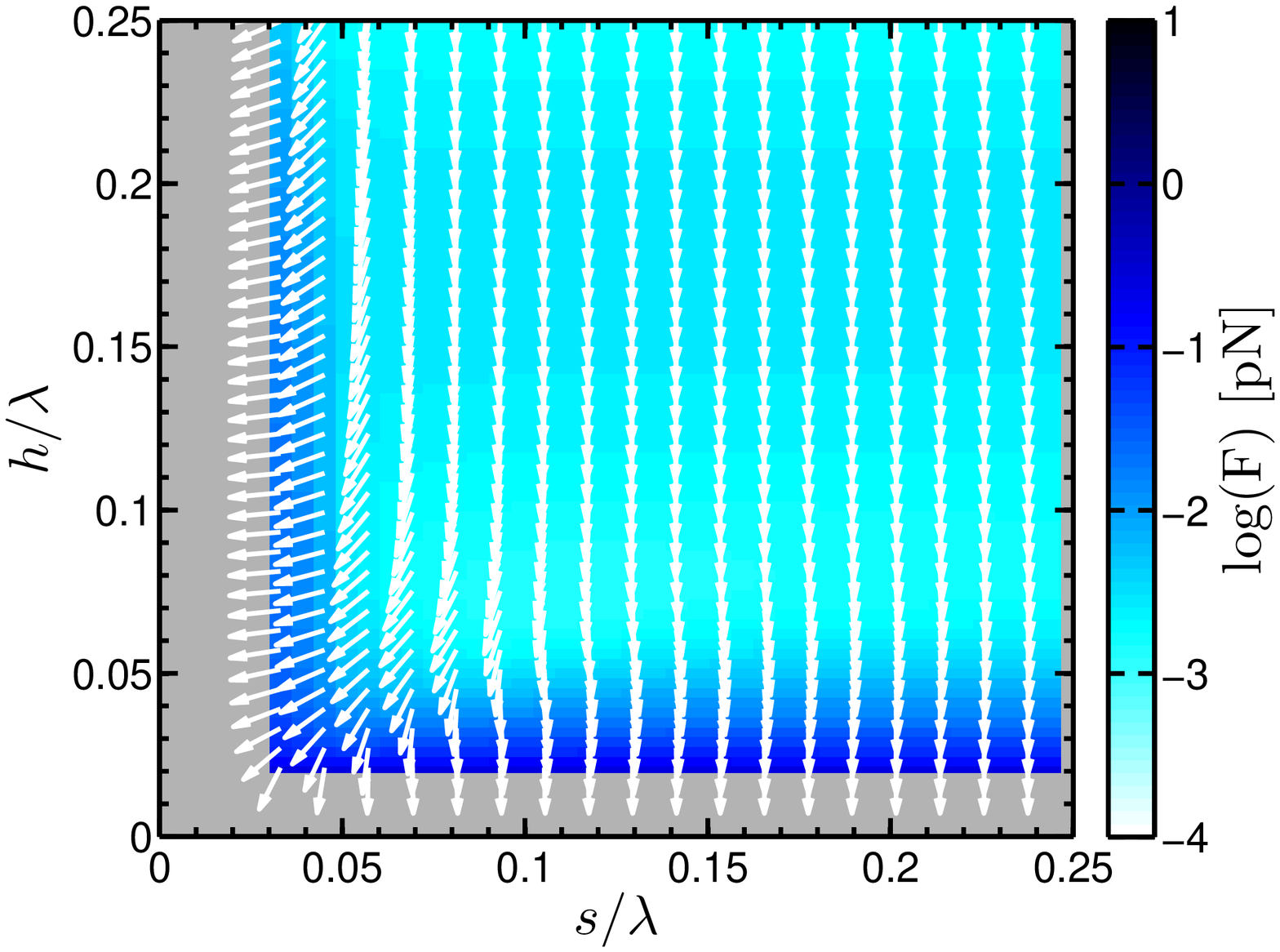}
}
\subfigure[\,Force lines on particle 2, at region-`iii' on Fig.~\ref{fig:GF_freq1}]{
\includegraphics[scale=0.4]{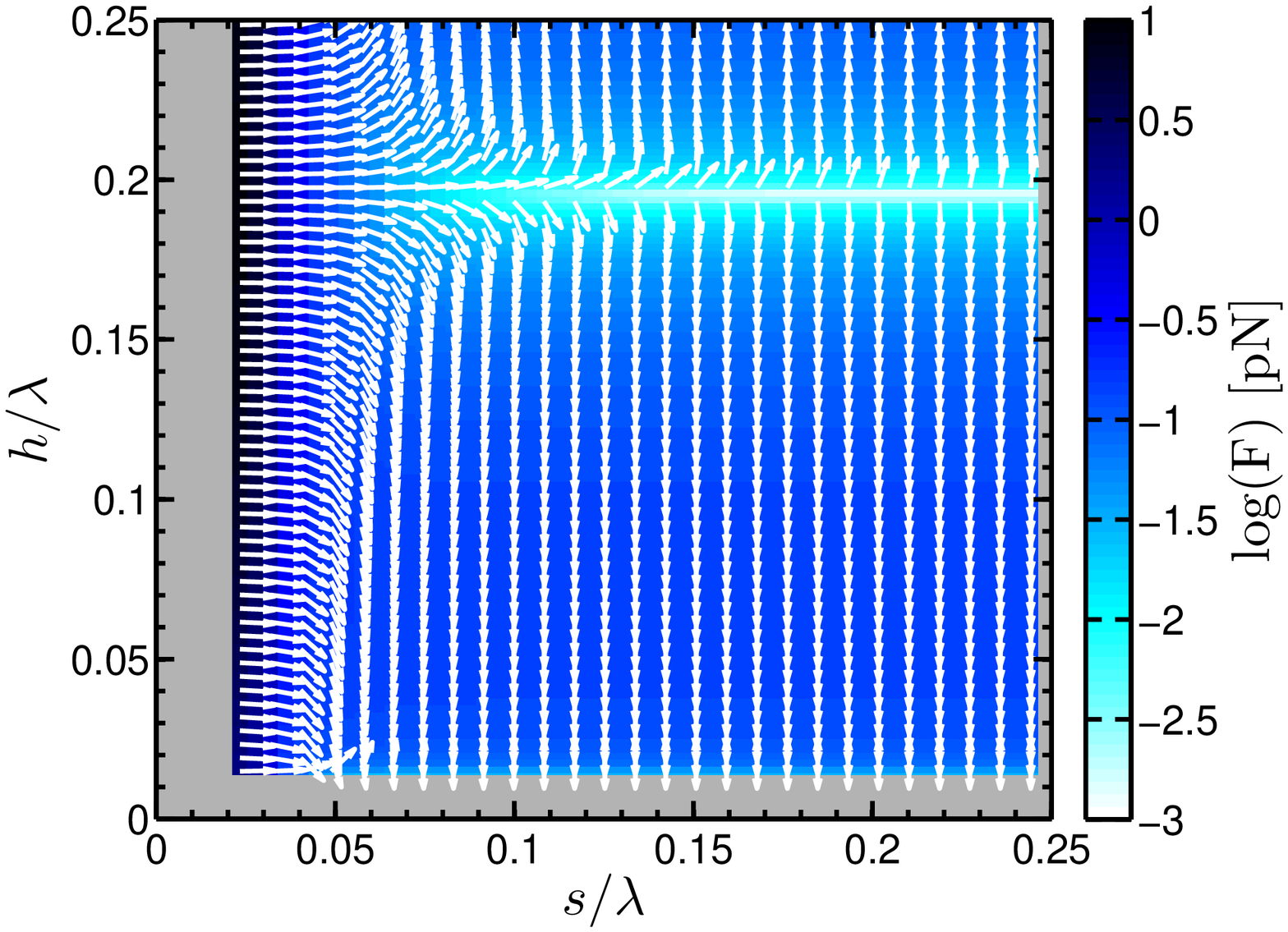}
}
\hfill
\subfigure[\,Force lines on particle 2, using at region-`iv'  on Fig.~\ref{fig:GF_freq1}]{
\includegraphics[scale=0.4]{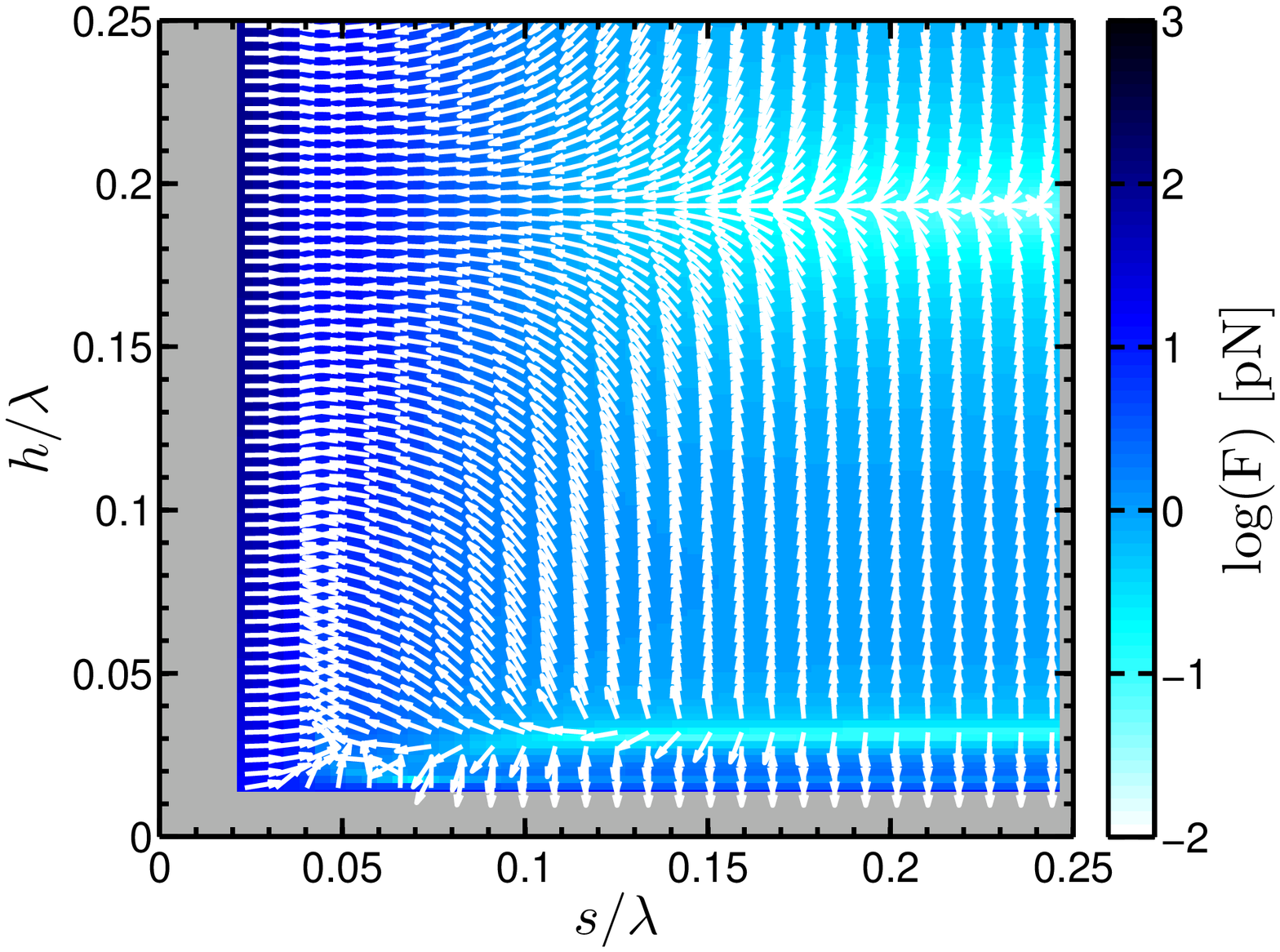}
}
\vspace{-0.2cm}
\caption{ (Color online)
{\em Silver half-space system showing the coupled nano-particle forces.}
(a)
Two dimensional graphs showing inter-particle forces as a function of height
and particle separation; the particles are illuminated from above with a plane wave polarized in the $y$ direction at the SPP frequency of the silver half-space and with the NP LSP tuned to be resonant with the SPP frequency. Colorbar indicates magnitude in log scale and arrows indicate directionality of the forces. One particle remains at $x_1=y_1=0$ and the second particle is moved in the $y$ direction. The two particles are then both varied in the $z$ direction such that they are always in the same plane. The arrows describe the force on the second particle which is not at the origin.
(b) As in (a), but now the NP is far red detuned from the SPP frequency to a LSP resonance of $\omega = 2.63$ eV.
 (c) As in (a), but the incident field is
at a frequency far detuned from the SPP frequency,  $\omega = 2.63$ eV. The NP LSP frequency of the particle is resonant with the SPP frequency.
(d) As in (c),  but with the NP ia also far red detuned to having a LSP resonance at $\omega_{LSP} = 2.63$ eV.
 \label{fig:Force1}}
\end{figure*}

It is also useful to examine the LDOS as a function of frequency for fixed particle position, as is shown in Figs.~\ref{fig:GF_freq1}(a-b) for different NP detunings (see Fig. \ref{fig:index_silver} and Eqn. \ref{eq:drude}). In both figures, the dashed lines correspond to $\rho^{(0)}$, the chain lines correspond to $\rho^{(1)}$, and the
 solid lines correspond to $\rho^{(2)}$. For simplicity we only focus on the imaginary part to examine the effects of the interactions.
  The particles both have their height fixed at $h= 0.05\,\lambda_{SPP}$, and their separation is $s=0.05\,\lambda_{SPP}$. In Fig.~\ref{fig:GF_freq1}(a) we consider a NP that is detuned by $\Delta \omega = 1.1$ eV,  and in Fig.~\ref{fig:GF_freq1}(b) the NP is on resonance with the SPP.
In Fig.~\ref{fig:GF_freq1}(a), the SPP is visible at $\omega = 3.73$ eV  in the LDOS when the particles are far detuned from this resonance, but the particles have a negligible effect on the SPP. Additionally, we see the resonances of the particles interacting and
produce a doublet feature caused by photon exchange effects.
 As the LSP resonances are moved towards the SPP resonance, the high frequency coupled LSP peak merges into the SPP resonance and acts to broaden it as well as detune it. The NLDOS behavior (not shown) mirrors the effects seen here where the NPs strongly renormalize the NLDOS in the regime of the NP LSP regardless of where the LSP is with respect to the SPP.
Similar effects for the LDOS and the NLDOS are seen in cavity-QED systems where the non-perturbative coupling between atoms or quantum dots causes additional photon exchange oscillations on top
of the vacuum Rabi oscillations~\cite{Yao-OEXP-17-11505} (the latter occur in systems with suitably small dissipation).

With the LDOS and NLDOS calculations acting as a guide, we can now examine the light-induced force calculations for the geometry shown in Fig.~\ref{fig:scheme} and described above. Four excitation regimes
of interest shown in Figs.~\ref{fig:Force1}(a)-(d), corresponding to the regions highlighted in Figs.~\ref{fig:GF_freq1}(a)-(b). We plot in log scale the intensity and use arrows to indicate the direction of the force. In Fig.~\ref{fig:Force1}(a) we illuminate at the SPP frequency and tune the LSP resonance of the particle to be at the same value. The particle separations and heights are both varied up to $0.25\,\lambda$ ($=83$~nm). For an inter-particle separation greater than $0.15\,\lambda$, the particles cannot feel each other except when the magnitude of the force is much less than 1~pN, which happens at a height of $\approx \, 0.125 \lambda$ and is due to the single particle interaction with the surface. The particles would thus be pushed away and then trapped in stationary positions $\approx\, 0.125 \lambda$ above the surface and at a separation of $0.185\,\lambda$. Interestingly, as the particles get closer to each other their interaction can still be negligible compared to the particle-surface interaction if their height is smaller than their inter-particle separation; this is caused by  quenching which reduces the transfer of radiation between the two
NPs. If the inter-particle separation is
 sufficiently close, and greater than their height above the surface, then the particles strongly
 optically couple to each other and to the half-space -- as can be seen by the fact that force still varies as the height of the particle varies.
The vector force topology seen in this graph manifests itself through the coupling between the systems constituents as their separation varies. This dynamic coupling is very similar to the self-induced
{\em back-action} demonstrated by Juan {\em et al.}~\onlinecite{Juan-NaturePhysics-5-915}.

In Fig.~\ref{fig:Force1}(b), we consider a similar excitation scenario as in Fig.~\ref{fig:Force1}(a), except that the LSP of the NP is far red detuned, with $\omega_{LSP} = 2.63$ eV. In this case, we notice that the silver half-space dominates the response and the particles are continually drawn to the surface unless $s <0.15\,\lambda$. When $s<0.15\,\lambda$ the particles can couple to each other and are drawn together, however the effects of the surface seem to be negligible for $h>0.1\,\lambda$.

In Fig.~\ref{fig:Force1}(c)
we tune our illumination to $\omega = 2.63$~eV, but keep our LSP resonant with the SPP; note that $0.25\,\lambda = $118\,nm. For particle separations greater than $0.08\lambda$, the effects of the surface dominate, and for separations below $0.08\,\lambda$, the inter-particle effects dominate  -- though they
still sensitively depend upon height.
Both Fig.~\ref{fig:Force1}(b) and (c) exhibit very weak inter-particle coupling and very weak particle-surface coupling, so that
 the perturbative expression for Eq.~(\ref{eq:En}) would hold. This is highlighted by the fact that the magnitude of the particle forces are much lower than when we illuminate on the LSP resonance.

For our final force example in Fig.~\ref{fig:Force1}(d), we examine the case when the LSP and the illumination are both far detuned
($\omega = 2.63$~eV, cf~$\omega_{SPP} = 3.73$~eV) from the SPP resonance.
We observe three useful coupling regimes:
$i$) When the particles are very close to the surface ($<0.05\,\lambda$), and the inter-particle separation is greater than $0.1\,\lambda$, we see that the surface completely dominates the forces and the particles are pulled towards the half-space. $ii$) When the particles are very close to each other ($<0.1\,\lambda$), then the inter-particle interaction dominates but this is again mediated by the half-space as there is a height dependence. $iii$) In the remaining region, we can see that both inter-particle coupling and surface-particle coupling is present where the half-space dominates when $h<s$ and particle-particle coupling is more dominant when $h>s$. This trend does not continue indefinitely as for $h>0.2\,\lambda$ we see the inter-particle forces become weaker for equivalent separations. The role of electromagnetic quenching is also minimal as we are so far from the ``lossy'' SPP resonance.

It is worth mentioning again, that the use of the gradient force instead of Eq.~(\ref{eq:F}) would predict an entirely different answer. Additional calculations (\emph{not shown}) show that for the case of Fig.~\ref{fig:Force1}(a), the gradient force topography is completely different, with an additional node along a vertical line at $s/\lambda \approx 0.16$ and no variation of force direction above $h/\lambda =0.1$. Thus using the gradient force for such a strongly perturbed system
is generally not valid.

\section{Negative Index Material Slab Waveguide\label{sec:metamaterial}}

\begin{figure}[t]
\includegraphics[width=3.25in]{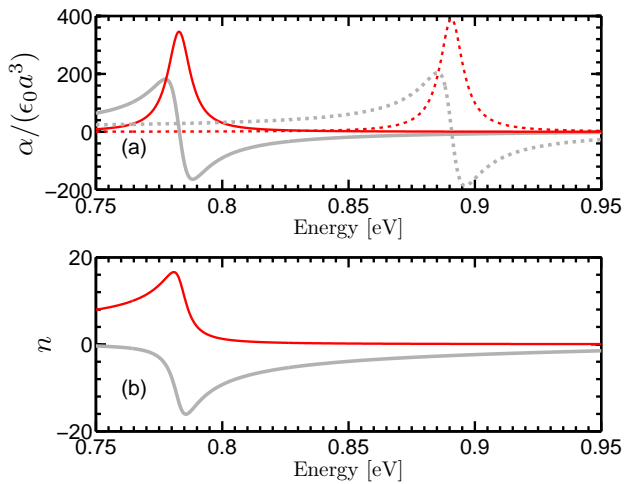}
\caption{(Color online) Particle and NIM slab response functions. For both the NIM and the NP permittivity we use the Drude model [Eq.~(\ref{eq:drude})], and for the NIM we model the permeability with a Lorentzian [Eq.~(\ref{eq:lorentz})]. (a) Bare polarizability [Eq.~(\ref{eq:alpha0})] of a nanoparticle with the LSP resonance tuned to the SLM of the 280 nm thick metamaterial slab, $\omega_{SLM} = \omega_{LSP} = 0.78$ eV (solid lines) ($\omega_{pe} = 2.22$ eV), and with the LSP resonance tuned to be off resonant with the LSP of the 280 nm metamaterial slab, $\omega_{LSP} = 0.89$ eV (dashed lines) ($\omega_{pe} = 2.53$ eV). In both cases $\gamma = 11$ meV. Gray-light curves indicates real parts and red-dark curves indicates imaginary parts. (b) Refractive index of the slab, where the gray-light curve indicates real part and the red-dark curve indicates imaginary part. Parameters are $\varepsilon_r=1$, $\omega_{pe} = 2.03$ eV, $\gamma = 8.3$ meV, $\omega_{pm} 0.69$ eV and $\omega_0 = 0.78$ eV.
\label{fig:index_mm}}
\end{figure}

We next examine a $280$-nm NIM metamaterial slab which supports a negative index in the frequency region $\omega = 0.78-0.92$~eV.
 The relevant NIM  and NP response functions are shown in Fig.~\ref{fig:index_mm}. The possible benefits of using metamaterials is the ability to tune the material properties by engineering the constituents of the unit cell. Negative index metamaterials can be produced with very low loss in the microwave regime, however scaling to the {\em visible} has proven to be quite a challenge as the materials become very lossy \cite{Boltassevaa-META-2-1}, though
 continued progress is being made with new designs~\cite{StanleyP.Burgos2010}. We will use NIM parameters that are close to experimental state-of-the-art for communications wavelengths, yet still have a respectable figure-of-merit: $FOM=|{\rm Re}(n)|/ {\rm Im}(n)$. Reported figure-of-merits are $FOM=2.0$ at 1.8~$\mu$m\cite{Zhang:06} and $FOM=0.5$ at 780~nm \cite{Dolling:07};  for our calculations, $FOM\approx1.0$ at $\omega = 0.78$~eV.
The permittivity is given by Eq.~(\ref{eq:drude}) with $\varepsilon_r=1$, $\omega_{pe} = 2.03$~eV and $\gamma = 8.3 $~meV. The permeability is given by the Lorentz model~\cite{Reza-NAT-455-312},
\begin{equation}
\label{eq:lorentz}
 \mu \left(\omega\right) = 1 + \frac{\omega_{pm}^2}{\omega_0^2-\omega^2 - i\omega\gamma},
\end{equation}
with the magnetic plasma frequency $\omega_{pm} = 0.69$~eV and the atomic resonance frequency $\omega_0 = 0.78$~eV.

Detailed descriptions of the exact corresponding {\em complex} band structure and Purcell effect are given by Yao {\em et al.}~\cite{Yao-PhysicalReviewB-80-195106} for the same parameters as given above. Here, we
briefly point out a few features of interest for this study.
At $\omega_0$, the dispersion curves of all of the leaky slow light modes (SLMs) of the system converge at a single frequency, $\omega_{SLM}=\omega_0$; thus particles near the slab can couple to many different slow light modes at this frequency. The slow light frequency regime gives rise to an enhancement in the LDOS and correspondingly to an increased Purcell effect.
 An increase in the LDOS is also seen near the SPP modes of the metallic surfaces, but the effects of quenching in metallic systems reduces the amount of light that escapes to the far field and the typical propagation distances of SPPs are limited by the material loss. However, slow light modes could, in principle, propagate for much longer distances. Also, the typical scaling laws associated with the quasi-static approximation are not reached, even very close to the slab~\cite{Yao-PhysicalReviewB-80-195106} (because of the strong magnetic resonance). Thus for our study, we are never really in the quasi-static regime above a NIM slab and we must consider retardation effects.
It is also worth noting that NIM slabs support both TE (transverse electric)
and TM  (transverse magnetic) SPPs, which is in contrast to metallic surfaces that only support TM SPPs. The SPP modes in NIMs will not be discussed here,
as their general properties are similar to the SPP of metals and at higher frequencies ($\omega_{SPP}^{TE} = 0.92$~eV and $\omega_{SPP}^{TM} = 1.43$~eV) \cite{Yao-PhysicalReviewB-80-195106}.

\begin{figure}[t]
\centering
\subfigure[\,LDOS]{
\includegraphics[scale=0.4]{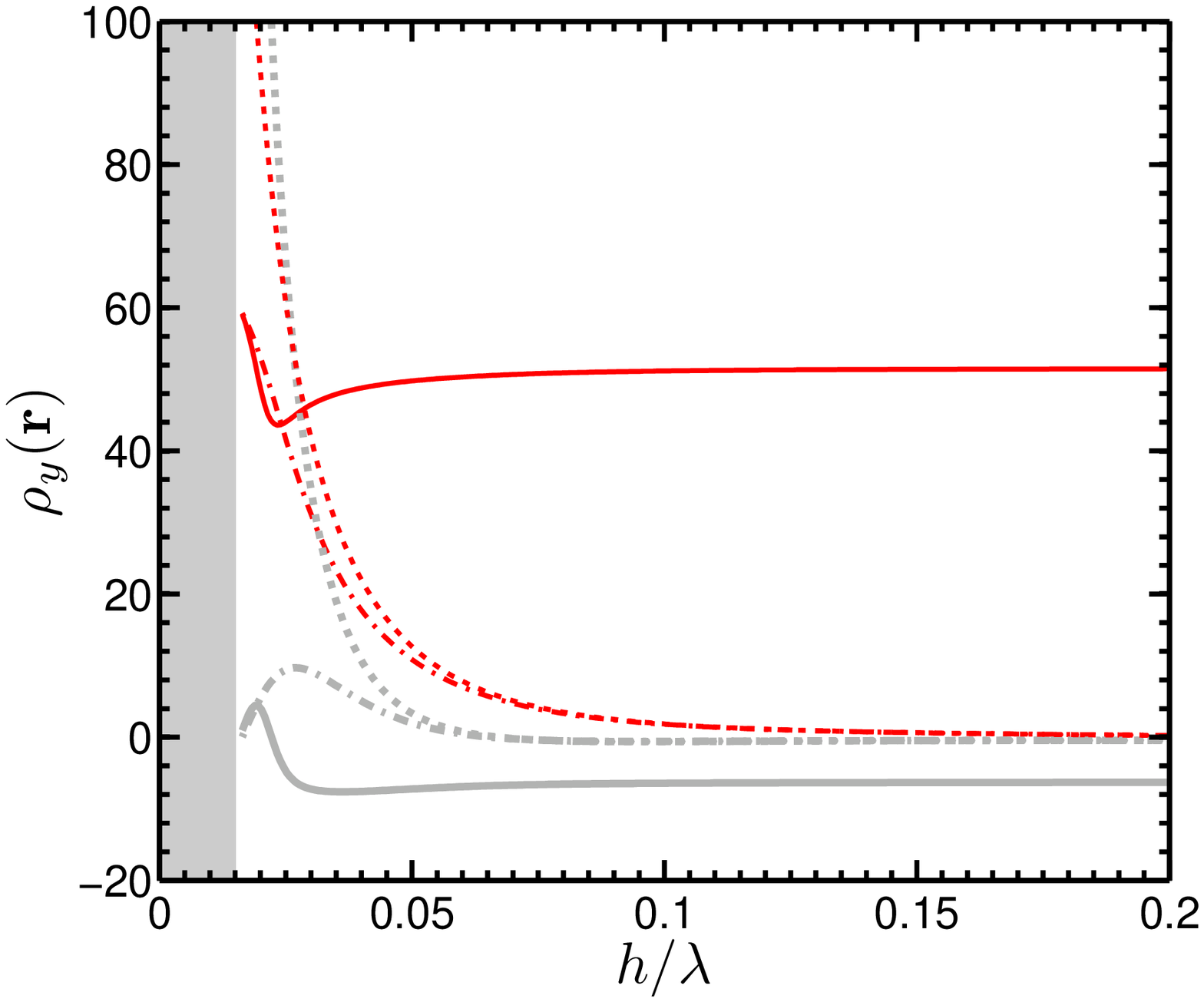}
}
\hfill
\subfigure[\,NLDOS]{
\includegraphics[scale=0.4]{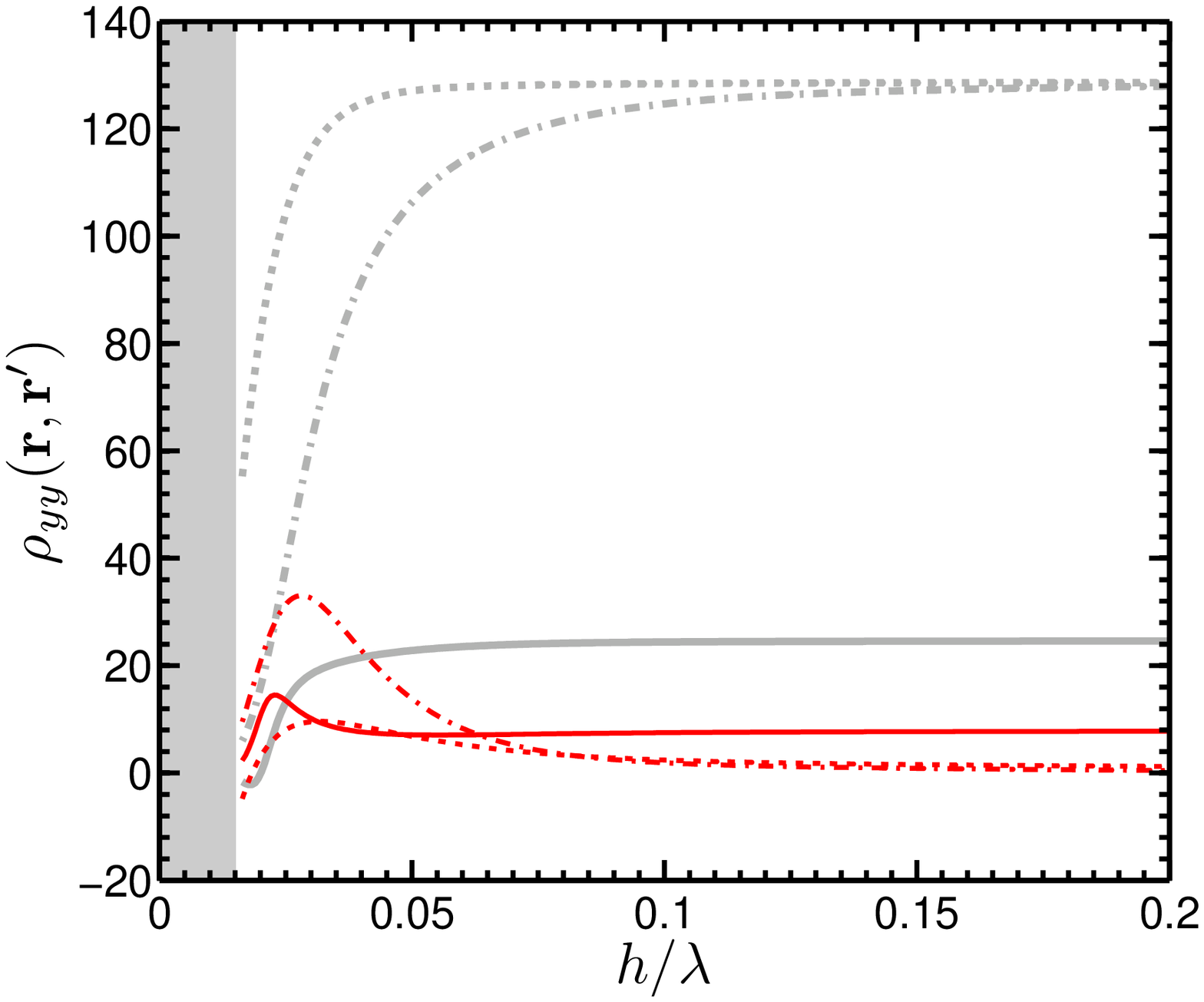}
}
\vspace{-0.2cm}
\caption{ (Color online)
(a) Complex LDOS, $\rho^{(N)}_{y}\left(\mathbf{r}', \omega\right)$, and (b) NLDOS, $\rho^{(N)}_{yy}\left(\mathbf{r},\mathbf{r}', \omega\right)$, for a 280~nm thick metamaterial slab as a function of height at $\omega = \omega_{SLM}$. For the LDOS $\mathbf{r} = \left(0,0,h\right)$, and for the NLDOS $\mathbf{r}' = \left(0,0,h\right)$, $\mathbf{r} = \left(0,0.05\lambda,h\right)$. Gray-light curves represent the real part and red-dark curves represent the imaginary part. The dashed line has no particles [$\rho^{(0)}_{y}\left(\mathbf{r};\omega\right)$, $\rho^{(0)}_{yy}\left(\mathbf{r},\mathbf{r}';\omega\right)$], the chain line has one particle located at the position the LDOS/NLDOS is calculated [$\rho^{(1)}_{y}\left(\mathbf{r};\omega\right)$, $\rho^{(1)}_{yy}\left(\mathbf{r},\mathbf{r}';\omega\right)$ and $\mathbf{r}_1 = \left(0,0,h\right)$], and the solid line has two particles; one where the LDOS/NLDOS is calculated at the other at the same height as the first but separated by $0.05 \lambda$ in the $y$ direction [$\rho^{(2)}_{y}\left(\mathbf{r};\omega\right)$, $\rho^{(2)}_{yy}\left(\mathbf{r},\mathbf{r}';\omega\right)$, $\mathbf{r}_1 = \left(0,0,h\right)$, and $\mathbf{r}_2 = \left(0,0.05\lambda,h\right)$]. Gray shaded region indicates region where particles would overlap the surface.
 \label{fig:GF_mm}}
\end{figure}

 For the metamaterial system, we consider a particle with a scaled radius of $0.015 \lambda_{SLM} = 24$~nm, and we tune our NP to be in the frequency regime of our peak LDOS associated with the SLMs (at $\omega_0$, see Fig. \ref{fig:index_mm}); practically, such tuning may be achieved,
 e.g., by using nano-shell structures~\cite{doi:10.1021/nn102035q}. Additionally, we reduce the NP damping rate to  $\gamma = 11$~meV to examine the SLM features which would otherwise be obscured.
For the metamaterial slab, we expect large enhancements of LDOS at the slow-light modes frequency similar to
Ref.~[\onlinecite{Yao-PhysicalReviewB-80-195106}], however it is not obvious what the inter-particle coupling effects will be, nor the role of inter-particle
coupling from the waveguide modes. Similar to the metal half-space case  [Figs.~\ref{fig:GF_silver} (a-b)],
we first examine the LDOS and the NLDOS in Figs.~\ref{fig:GF_mm}(a-b) as a function of height.
In Fig.~\ref{fig:GF_mm}(a), the real (gray-light curve) and imaginary parts (red-dark curve) of the LDOS again diverge as the slab is approached [Eq.~(\ref{eq:mult_G_tot_slab_QS})] when no particles are in the system, however there is a change in sign of the real part compared to the metallic case, indicating the Lamb shift would be a red shift instead of a blue shift. The imaginary part (Purcell factor) reaches a value of 100 at $0.02\lambda_{SLM} = 31.7$~nm compared to the metallic case which reaches a value of 100 at $0.05\,\lambda_{SPP} = 16.6$~nm. Thus the metamaterial gives an equivalent enhancement at twice the distance in absolute units. Introducing a particle at the location of the LDOS [$\mathbf{r_1} = \mathbf{r} = \left(0,0,h\right)$] renormalizes both the real and imaginary parts of the LDOS at small $h/\lambda$ and removes some of the
divergence behavior for small distances close to the slab -- similar to the metallic case. We also
 see that the maximum of the real part is no longer located closest to the surface. The addition of the second particle [$\mathbf{r}_2 = \left(0,0.05\lambda,h\right)$] increases the imaginary part to a constant for $h>0.08\,\lambda$ to almost exactly the same value as for the metallic case which is due to the fact that we are in the quasi-static limit for the homogenous interaction between the particles and the slab no longer plays a role. The real part dips slightly below zero indicating that there can be either a blue or a red shift depending on the height of the particles and stays below zero for $h>0.025\,\lambda$.

For the NLDOS [Fig.~\ref{fig:GF_mm}(b)], the real part (gray-light curve) follows a very similar trend as in the metallic case, where the zero particle case (dashed curve) is reduced as the slab is approached but is finite. Including the first particle (chain) drastically decreases the real part and thus the virtual photon exchange and the second particle (solid) further reduces it. At closest approach the real part is small but still greater in magnitude to the metallic case by a factor of 20. The imaginary part with no particles (red-dark dashed curve) qualitatively follows the metallic case however when a particle is included in the system (red-dark  chain curve), instead of reducing the real photon transfer there is an increase. The inclusion of the second particle (solid curve) reduces the effect again but we still are able to increase coupling between the particles compared to the metallic case. This transfer can be further increased as it crucially depends on the metamaterial loss used in the effective permittivity and permeability. Obtaining lower losses is possible by improving metamaterial fabrication techniques which would result in less lossy slow-light propagation modes.

\begin{figure}[t]
\includegraphics[width=3.25in]{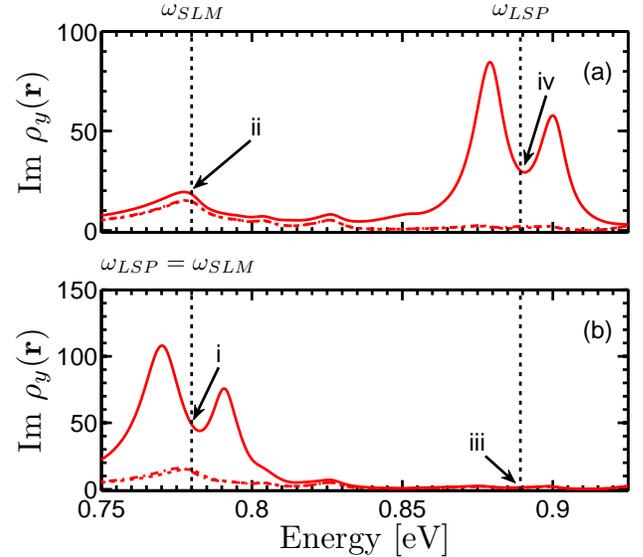}
\caption{(Color online) Im[$\rho^{(N)}_{y}\left(\mathbf{r}, \omega\right)$] for a 280~nm thick metamaterial slab as a function of frequency at the position $\mathbf{r} = \left(0,0,0.05\lambda_{SLM}\right)$.
  The dashed line has no particles, Im[$\rho^{(0)}_{y}\left(\mathbf{r};\omega\right)$],
 the chain line has one particle, Im[$\rho^{(1)}_{y}\left(\mathbf{r};\omega\right)$], and
 the solid line has two particles, Im[$\rho^{(2)}_{y}\left(\mathbf{r};\omega\right)$], at locations
 $\mathbf{r}_1 = \left(0,0,0.05\,\lambda_{SLM}\right)$, and $\mathbf{r}_2 = \left(0,0.05\lambda_{SLM},0.05\lambda_{SLM}\right)$. (a) The particle LSPs are blue detuned by $\Delta \omega = 0.11$~eV compared to the slab SLM frequency (see Fig. \ref{fig:index_mm}). (b) The particle LSPs are at the slab SLM frequency. Arrows indicate the particular scenarios we are examining in force graphs: `i' -- Fig. \ref{fig:Force5}(a), `ii' -- Fig. \ref{fig:Force5}(b),
`iii' -- Fig. \ref{fig:Force5}(c) and `iv' -- Fig.~\ref{fig:Force5}(d).
\label{fig:GF_freq1_mm}}
\end{figure}

A comparison of the LDOS in terms of frequency for the NIM slab is shown in
 Fig.~\ref{fig:GF_freq1_mm} for different NP detunings. Again, the particles both have their height fixed at $z = z' = 0.05\,\lambda_{SLM} =79$~nm, and their separation is $0.05\,\lambda_{SLM}$. In Fig.~\ref{fig:GF_freq1_mm}(a) we consider a NP that is blue detuned by $\Delta \omega = 0.11$~eV, and Fig.~\ref{fig:GF_freq1}(b), the NP is on resonance with the slab SLMs.
When the NP is detuned from the slow-light modes (non-resonant case), we see that there is still a large enhancement of the LDOS at the NP resonance compared with the zero particle case, but this is essentially the homogeneous space coupling due to the particles and is only slightly altered by the presence of the slab. When the NP is tuned to the $\omega_{SLM}$ resonance there is a greater enhancement than in free space but the coupling is dominated by inter-particle interactions. If we include only one particle, it is evident that the inter-particle effects dominate the spectrum as the LDOS more closely follows the zero particle case.

\begin{figure*}[htb]
\centering
\subfigure[\,Force lines on particle 2, at region-`i' on Fig.~\ref{fig:GF_freq1_mm}]{
\includegraphics[scale=0.4]{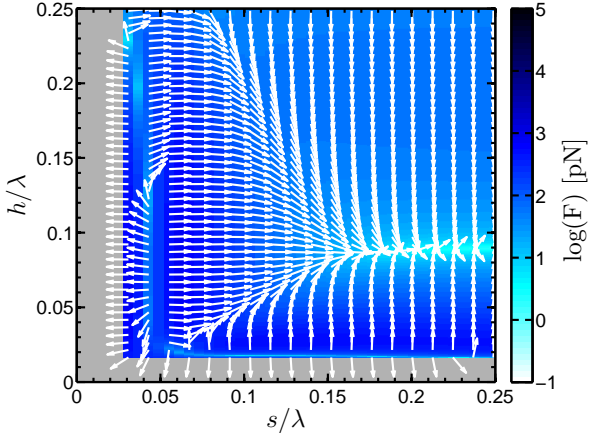}
}
\hfill
\subfigure[\,Force lines on particle 2, at region-`ii' on Fig.~\ref{fig:GF_freq1_mm}]{
\includegraphics[scale=0.4]{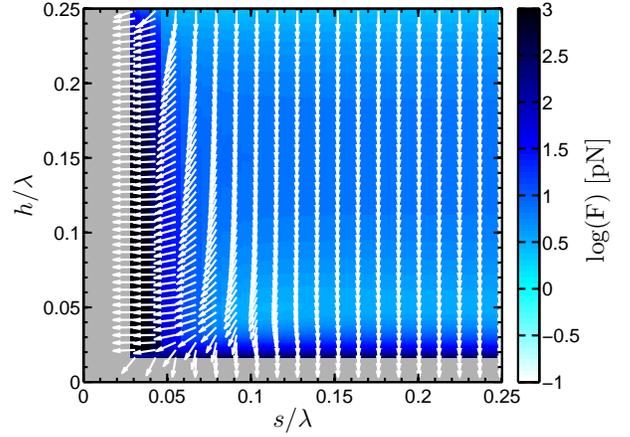}
}
\subfigure[\,Force lines on particle 2, at region-`iii' on Fig.~\ref{fig:GF_freq1_mm}]{
\includegraphics[scale=0.4]{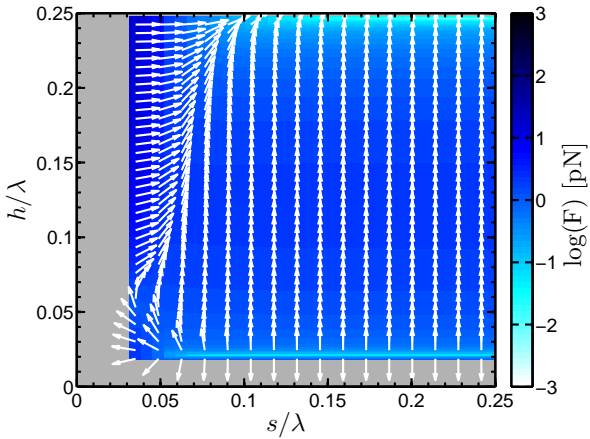}
}
\hfill
\subfigure[\,Force lines on particle 2, using at region-`iv'  on Fig.~\ref{fig:GF_freq1_mm}]{
\includegraphics[scale=0.4]{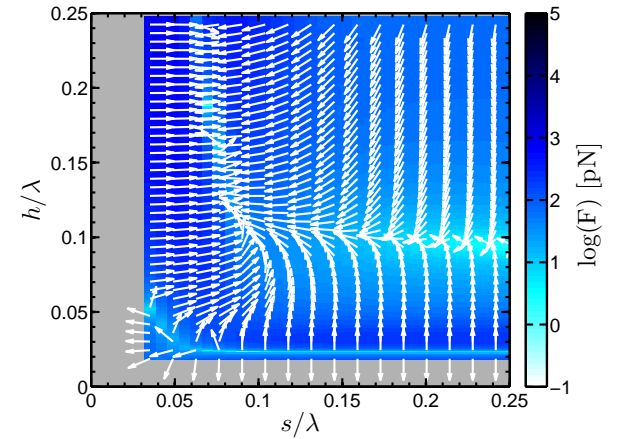}
}
\vspace{-0.2cm}
\caption{ (Color online)
{\em Metamaterial slab system showing the coupled nano-particle forces.}
(a)
Two dimensional graphs showing inter-particle forces as a function of height
and particle separation; the particles are illuminated from above with a plane wave polarized in the $y$ direction at the SLM frequency of the slab ($\omega_{SLM} =0.78$~eV). The NP LSP is also tuned to be resonant with the SLM frequency. Here one particle remains at $x_1=y_1=0$ and the second particle is moved in the $y$ direction. The two particles are then both varied in the $z$ direction such that they are always in the same plane. The arrows describe the force on the second particle which is not at the origin.
(b) As in (a), but now the NP is blue detuned
from the SLM frequency to a LSP resonance of $\omega = 0.89$~eV.
 (c) As in (a), but the incident field is
at  detuned to $\omega = \omega_{LSP} =0.89$ eV.
(d) As in (c),  but with the NP ia also detuned to having a LSP resonance at $\omega = 0.89$~eV.
 \label{fig:Force5}}
\end{figure*}

To illustrate the effect on light-induced forces, we  again consider four different cases that are highlighted in Figs.~\ref{fig:GF_freq1_mm}(a-d). We  illuminate with a plane wave at the slow-light resonance frequency of the metamaterial slab  ($\omega_{SLM} =0.78$~eV), first with the NP on-resonance [Fig.~\ref{fig:Force5} (a)], and then with the NP off-resonance [Fig.~\ref{fig:Force5}(b)]. We then illuminate off resonance ($\omega = 0.89$~eV) tuning the nanoparticle to be on-resonance with the
 slow-light modes [Fig.~\ref{fig:Force5}(c)], and then to be off-resonant and at the same frequency of the illumination [Fig.~\ref{fig:Force5}(d)]. All figures show the log of the magnitude of the force in intensity scale and arrows indicate directionality.

When the NP and the slow-light modes are both on-resonance with the incident radiation [Fig.~\ref{fig:Force5}(a)], we see a very similar situation as when the NP was on resonance with the SPP [Fig.~\ref{fig:Force1}(a)]. For $s<0.2\,\lambda$ there is a division along the line $s\approx 2\,h$ where below this line the slab dominates the forces, above this line the inter-particle interaction dominates the forces and around which we see a combination of the two. As the height gets above $0.1\,\lambda$ we see that this does not continue indefinitely and the inter-particle interaction becomes weaker and shorter ranged as the slab no longer enhances the coupling between the two. Particles that have a small initial separation will be pushed away from each other and to a height of $\approx 0.08\,\lambda$. Note the forces here are an order of magnitude greater than in the metallic case, which is mostly due to the polarizability scaling with the particle size but these could in principle be further tuned by improving the loss in these structures.

When the NP is tuned off-resonance from both $\omega_{SLM}$ and the incident frequency [Fig.~\ref{fig:Force5}(b)], the long range coupling is lost compared to Fig.~\ref{fig:Force5}(a). For $s>0.12\lambda$, the force is dominated by slab interactions and are continually drawn to the surface of the slab.
For $s<0.1\,\lambda$, and $h<0.1\,\lambda$ we see that the particles and the slab are all interacting which results in the particles being pulled towards the slab and together however the interaction range is short. When $h>0.1\,\lambda$ the particles essentially only interact with each other and are mostly drawn together.

Figures \ref{fig:Force5}(c-d) show light-induced force calculations with the incident radiation detuned from the
 slow-light mode frequency to $\omega = 0.89$~eV. In Fig.~\ref{fig:Force5}(c) the NP LSP is tuned to be resonant with the SLMs and off-resonant with the radiation. For $s>0.1\,\lambda$ the particles are unaffected by each other and are repelled from the slab, except when they are almost touching. For $s<0.1\lambda$ the slab dominates over the inter-particle interaction for $h<0.1\,\lambda$ which is in contrast to Fig.~\ref{fig:Force5}(b). For $h>0.1\,\lambda$ the slab interaction weakens and the particles begin to interact and repel each other.

\begin{figure}[t]
\includegraphics[width=3.25in]{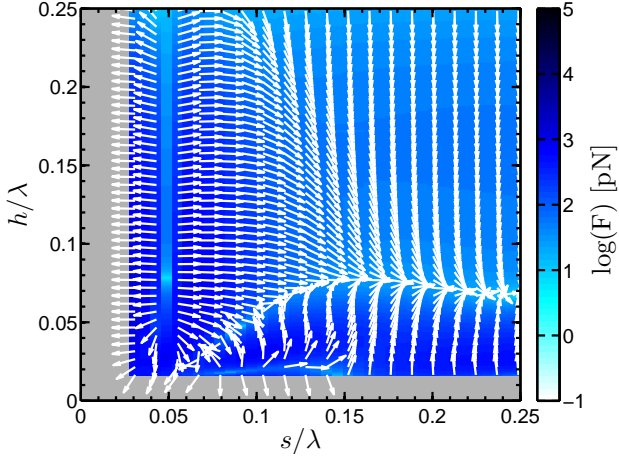}
\caption{ (Color online)
As in Fig. \ref{fig:Force5}(a) but with a low msterial loss, ($\gamma' = \gamma/10$)
 \label{fig:Force_LL}}
\end{figure}

\begin{figure}[th]
\centering
\subfigure[\,NLDOS for a nominal material loss, $\gamma = 8.3$~meV]{
\includegraphics[scale=0.4]{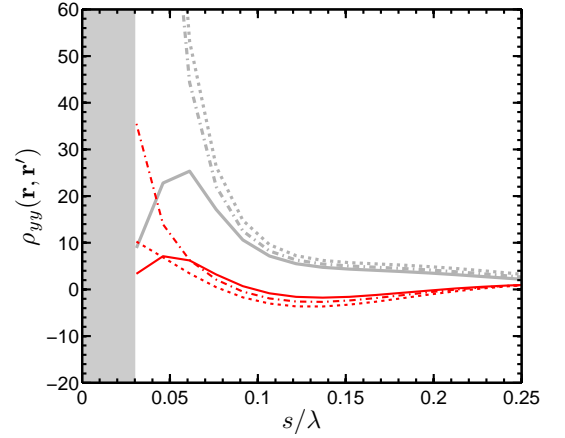}
}
\hfill
\subfigure[\,NLDOS for a lower material loss,  $\gamma = 0.83$~meV]{
\includegraphics[scale=0.4]{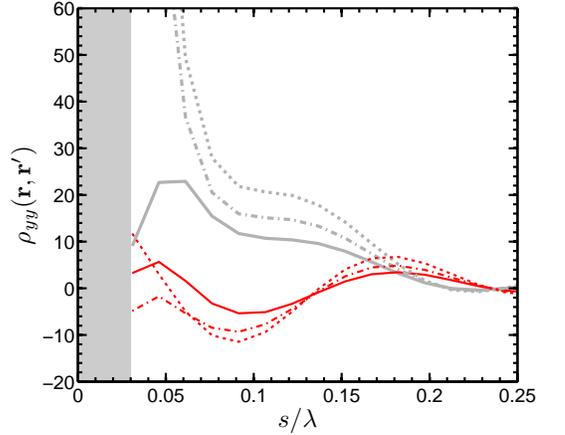}
}
\vspace{-0.2cm}
\caption{ (Color online)
Complex NLDOS, $\rho^{(N)}_{yy}\left(\mathbf{r},\mathbf{r}', \omega\right)$, for a 280~nm thick and metamaterial slab with regular loss (a) or with low-loss (b) as a function of separation [$\mathbf{r}' = \left(0,0,0.05 \lambda\right)$, $\mathbf{r} = \left(0,s,0.05 \lambda\right)$] at $\omega = \omega_{SLM}$. Gray-light curves represent the real part and red-dark curves represent the imaginary part. The dashed line has no particles [$\rho^{(0)}_{yy}\left(\mathbf{r},\mathbf{r}';\omega\right)$], the chain line has one particle located at the position the NLDOS is calculated [$\rho^{(1)}_{yy}\left(\mathbf{r},\mathbf{r}';\omega\right)$ and $\mathbf{r}_1 = \left(0,0,0.05 \lambda\right)$], and the solid line has two particles; one where the NLDOS is calculated at the other at the same height as the first but separated along the $y$ direction [$\rho^{(2)}_{y}\left(\mathbf{r},\mathbf{r}';\omega\right)$, $\mathbf{r}_1 = \left(0,0,0.05 \lambda \right)$, and $\mathbf{r}_2 = \left(0,s,0.05 \lambda\right)$]. Gray shaded region indicates region where particles would overlap each other.
 \label{fig:extra5}}
\end{figure}

Next, we examine the case where both the radiation and the NP are detuned to $0.89$~eV and away from the SLMs, shown in Fig. \ref{fig:Force5}(d). We see that, similar to Fig.~\ref{fig:Force1}(d),  there are three different interaction regions, $i$) $h<0.08\,\lambda$ and $s>0.1\,\lambda$, $ii$) $h<0.08\,\lambda$ and $s<0.1\,\lambda$, $iii$) and $h>0.08\,\lambda$. In the first region, the slab dominates the force by pulling the particle when it is almost touching and pushing the particle away when it is slightly higher $h>0.02\,\lambda$ until the point where the vertical force becomes negligible and the particles are attracted to each other at $h\approx 0.08\,\lambda$. In region $ii$), the particles are causing a dramatic renormalization of the Green function which leads to very strong, position dependent particle interactions which are pushing the particles away from the slab and each other until they get to $h\approx0.1\,\lambda$, $s\approx0.1\,\lambda$. Finally, above $h=0.1\,\lambda$ the particle interaction dominates but is mediated by the height above the slab and causes the particles to essentially be repelled to a fixed separation of $s\approx0.08\,\lambda$, as the separation increases the slab starts to draw the particles towards it again however the particles will still be pulled towards $s\approx0.08\,\lambda$.

To investigate the influence of metamaterial loss on the inter-particle forces, we show the same scenario as Fig. \ref{fig:Force5} (a) where the LSP and radiation is resonant with the SLMs, but we now decrease the material loss by a factor of 10, thus
$\gamma = 0.83$~meV. We show the resulting force in Fig. \ref{fig:Force_LL}, and see a number of
 important differences. First, where there was once a fixed height at which the particles would be attracted to, this height now varies with inter-particle separation. Below this dividing line in the region where $h<0.08\,\lambda$ and $s>0.18\,\lambda$ there are still inter-particle forces where in the regular loss case these forces have since died away. Finally, in the regular loss case, as the particles are moved vertically, along the line $s = 0.04\,\lambda$ we see that the particles are attracted to each other close the slab, $h<0.125\,\lambda$, but are repulsive at higher distances. This is contrasted in the low loss case where there is a division at $s=0.045\,\lambda$, below which the particles are always attracted and above which the particles are always repelled.

To further examine how the loss alters the long-range coupling effects, we plot the NLDOS in Fig.~\ref{fig:extra5} for both regular and low loss metamaterial slabs at a fixed height, $h = 0.05\,\lambda$, and vary the separation between $\mathbf{r}' = \left(0,0,0.05\lambda\right)$ and $\mathbf{r} = \left(0,s,0.05\,\lambda\right)$. In both cases we see that the real part (gray-light curve) diverges at low $s$ when there are no particles in the system (dashed line), due to the homogeneous part of the
Green function. The addition of particles once more renormalizes these values.
We also see that in the low loss case, the real part plateaus between $s=0.08\,\lambda$ and $s=0.0125\,\lambda$, whereas in the nominal loss case this simply decays, so material loss has a large influence on the light-induced forces The imaginary part of the NLDOS (red-dark curves) varies slowly towards zero in the regular loss case, however we see that the imaginary part of the NLDOS in the low loss case oscillates around a value of -2, with a much larger amplitude. The increase in the NLDOS allows the particles to couple much farther in Fig.~\ref{fig:Force_LL} than in Fig.~\ref{fig:Force5} (a). Thus for decreasing material losses, the NPs can be
coupled over longer distances where this coupling is mediated by the slow light waveguide modes.

\section{Conclusions}
 \label{sec:conclusions}
 We have introduced a theoretical formalism to compute the Green function response
 of small particles within the vicinity of multi-layered geometries.
 We have applied this theory to  calculate the non-perturbative force interactions between
 two NPs in the vicinity of surface plasmon polariton modes for
 a metal half-space geometry, and in the vicinity of
 a slow-light  NIM waveguide. Both planar structures facilitate
 a large local density of states, and non-local photon interactions between
 the particles. We have found that both structures exhibit rich but similar force maps despite the different mechanisms for increasing the LDOS. When both particle and slab (metal and NIM) are on resonance with the incident illumination, the particles will be pushed away from each other and pushed to a fixed height above the slab (Figs.~\ref{fig:Force1}(a) and \ref{fig:Force5}(a)). Such an effect would aid in preventing the aggregation of NPs. When the particle and illumination are off resonance with the slab then the particles are pulled to a specific height and pulled towards each other up to a fixed distance (Figs.~\ref{fig:Force1}(d) and \ref{fig:Force5}(d)) which would enable the creation of dimers. In all the other cases the most likely scenarios are the particles being pulled towards the slab or the particles being pushed away from the slab.

 For metallic surfaces, the material parameters are largely fixed, limiting some of the engineering available to such structures, however metamaterials in principle have the ability to have their intrinsic parameters tuned by changing the basic unit cell. Such tunability will allow simplified geometries such as the planar structures to aid in the creation of long range optical forces for the trapping and localization of small NPs. The same structures also exhibit rich and fundamentally interesting QED phenomena offering
 applications for radiative decay engineering of embedded quantum light sources.

\begin{acknowledgments}

This work was supported by the National Sciences and
Engineering Research Council of Canada and the Canadian
Foundation for Innovation. We gratefully acknowledge Mark Patterson
for assistance with the multi-layered Green function calculations.
\end{acknowledgments}

\appendix
\section{Planar Green function \label{sec:slab_GF}}
The Green function above a planar structure can be written in terms of its angular spectrum~\cite{Novotny-Nano-Optics} which involves decomposing the wavevector $\mathbf{k} = k_x \mathbf{\hat{x}}+k_y \mathbf{\hat{y}}+k_z \mathbf{\hat{z}}$ into its various components and integrating over each contribution. The real-space homogeneous Green function
\begin{widetext}
\begin{equation}
\label{eq:mult_G_hom}
\begin{split}
 &\GFT_{\rm hom}\rrarg = -\frac{\mathbf{\hat{z}}\mathbf{\hat{z}}}{\varepsilon_B}\delta\left(\mathbf{R}\right)
+\frac{i\omega^2}{8\pi^2 c^2 \varepsilon_B} \int_{-\infty}^{\infty}\int_{-\infty}^{\infty} \fT_{\rm hom} e^{i k_x\left(x-x'\right)+i k_y\left(y-y'\right)+i k_z\left|z-z'\right|} {\rm d}k_x {\rm d}k_y ,
\end{split}
\end{equation}
where $k_B = \omega\sqrt{\varepsilon_B}/c$ is the wavevector in the background material, and the z-component is given by $k_z = \left(k_B^2 - k_x^2-k_y^2\right)^{1/2}$. The matrix $\fT_{\rm hom}$ is given by
\begin{equation}
 \fT_{\rm hom} = \frac{1}{k_z} \begin{bmatrix}
       k_B^2-k_x^2 & -k_x k_y  & \mp k_x k_z \\
       -k_x k_y & k_B^2-k_y^2  & \mp k_y k_z \\
       \mp k_x k_z & \mp k_y k_z & k_B^2-k_z^2
     \end{bmatrix} ,
\end{equation}
where the upper sign is used when $z>z'$ and the lower sign is used when $z<z'$. The scattered part of the Green function in a multilayer environment (no particles) can be written similarly in terms of $s$ and $p$ polarized contributions,
\begin{equation}
\label{eq:mult_G_scatt}
\begin{split}
 \GFT_{\rm scatt}\rrarg = \frac{i\omega^2}{8\pi^2 c^2 \varepsilon_B} \int\int_{-\infty}^{\infty} \left[\fT^s_{\rm scatt} +\fT^p_{\rm scatt}\right]e^{i k_x\left(x-x'\right)+i k_y\left(y-y'\right)+i k_z\left(z-z'\right)} {\rm d}k_x {\rm d}k_y,
\end{split}
\end{equation}
\begin{equation}
\begin{split}
 \fT^s_{\rm scatt}= \frac{r^s\left(k_x,k_y\right)}{k_z\left(k_x^2+k_y^2\right)} \begin{bmatrix}
       k_y^2 & -k_x k_y  & 0 \\
       -k_x k_y & k_x^2  & 0 \\
       0 & 0 & 0
     \end{bmatrix},
\end{split}
\end{equation}
\begin{equation}
\begin{split}
 \fT^p_{\rm scatt} = \frac{r^p\left(k_x,k_y\right)}{k_B^2\left(k_x^2+k_y^2\right)} \begin{bmatrix}
       k_B k_x^2 & k_x k_y k_z  & k_x \left(k_x^2+k_y^2\right) \\
       k_x k_y k_z & k_B k_y^2  & k_y \left(k_x^2+k_y^2\right) \\
       - k_x \left(k_x^2+k_y^2\right) & -k_y \left(k_x^2+k_y^2\right) & -\left(k_x^2+k_y^2\right)/k_z
     \end{bmatrix}.
\end{split}
\end{equation}
\end{widetext}
Here the matrices (or dyadics) $\fT^s_{\rm scatt}$ and $\fT^p_{\rm scatt}$ are given in terms of the reflection coefficients, $r^{s/p}$, for $s$ and $p$ polarization above the multilayer. For the three-layer geometry of a slab with height $h$ considered here, the upper, background layer having $\varepsilon_B = \varepsilon_1,\mu_B = \mu_1$, the middle layer having $\varepsilon_2,\mu_2$ and the lower layer having $\varepsilon_3,\mu_3$ these reflection coefficients are
\begin{equation}
 r^{s/p} = r^{s/p}_{12} + \frac{t^{s/p}_{12} t^{s/p}_{21} r^{s/p}_{23} e^{2i\beta h}}{1 -r^{s/p}_{21} r^{s/p}_{23} e^{2i\beta h}} ,
\end{equation}
where $\beta = \pm \left(\omega^2 \varepsilon_2 \mu_2  /c^2 - k_x^2 - k_y^2\right)^{1/2}$ is the $z$ component of the wavevector in the middle layer when ${\rm Re }\left(\omega^2 \varepsilon_2 \mu_2  /c^2\right)>{\rm Re }\left(k_x^2 + k_y^2\right)^{1/2}$. The sign of $\beta$ depends on whether or not the refractive index of the middle layer is positive (upper) or negative (lower) \cite{Dung-PRA-68-043816,Ramakrishna-OLETT-30-2626}. For ${\rm Re }\left(\omega^2 \varepsilon_2 \mu_2  /c^2\right)<{\rm Re }\left(k_x^2 + k_y^2\right)^{1/2}$, $\beta = i \left(k_x^2 + k_y^2 - \omega^2 \varepsilon_2 \mu_2 /c^2 \right)^{1/2}$ for both positive and negative index materials. The single-layer reflection and transmission coefficients are
\begin{eqnarray}
 r_{ij}^s = \frac{\mu_j k_{iz} - \mu_i k_{jz}}{\mu_j k_{iz} + \mu_i k_{jz}}, \quad \quad r_{ij}^p = \frac{\varepsilon_j k_{iz} - \varepsilon_i k_{jz}}{\varepsilon_j k_{iz} + \varepsilon_i k_{jz}}\\
 t_{ij}^s = \frac{2\mu_j k_{iz}}{\mu_j k_{iz} + \mu_i k_{jz}}, \quad \quad t_{ij}^p = \frac{2\varepsilon_j k_{iz}}{\varepsilon_j k_{iz} + \varepsilon_i k_{jz}}.
\end{eqnarray}

Solutions of Eqs.~(\ref{eq:mult_G_hom}) and (\ref{eq:mult_G_scatt}) for real dielectrics can be difficult due to poles close or along the path of integration, however this can be solved by numerically integrating around the poles in the complex plane which lie in a known region~\cite{Paulus-PRE-62-5797}. For lossy NIMs the poles along the integration path are found to be in the lower part of the complex plane, whereas for lossy positive index materials the poles are located in the upper half of the complex plane~\cite{Yao-PhysicalReviewB-80-195106}. The solution described by Paulus {\it et. al.}~\cite{Paulus-PRE-62-5797} is more complicated for materials which are able to support negative index modes and surface plasmons as the location of the poles in the complex plane are essentially given by the {\em complex} band structure of the material~\cite{PhysRevB.79.195414}. The largest contributions to the Green function are no longer confined to the region where ${\rm Re }\left(\omega^2 \varepsilon_2 \mu_2  /c^2\right)>{\rm Re }\left(k_x^2 + k_y^2\right)^{1/2}$ and careful attention must be paid to the integrand. This is trivial for a small number of calculations but can be cumbersome when many locations are required. As an example, for a two particle force calculation at a single point, the above calculations required 14 separate Green function calculations when employing the dipole approximation.

When considering the Green function in the quasi-static approximation, the homogeneous space Green function~\cite{Gay-Balmaz:01} is given by
\begin{equation}
\label{eq:mult_G_hom_QS}
\begin{split}
 &\GFT_{\rm hom,QS}\rrarg = \frac{1}{4 \pi \varepsilon_B  R^3}\left( \frac{3\mathbf{RR}}{R^2} -\IT \right) ,
\end{split}
\end{equation}
where $\IT$ is the unit dyadic (diagonal terms are unity and non-diagonal terms are zero).
The total Green function above a half-space in the quasi-static approximation involves the direct contribution from Eq.~(\ref{eq:mult_G_hom_QS}) as well as the scattered contribution from an image source located beneath the surface \cite{GayBalmaz200037},
\begin{equation}
\label{eq:mult_G_tot_slab_QS}
\begin{split}
 &\GFT_{\rm QS}\rrarg = \GFT_{\rm hom,QS}\rrarg \mp \frac{\varepsilon_2 - \varepsilon_1}{\varepsilon_2 + \varepsilon_1} \GFT_{\rm hom,QS} \left(\br,\br''\right).
\end{split}
\end{equation}
Here the minus sign is for x/y directed dipoles and the plus sign is for z directed dipoles. The location of the image charge is given by $\br''$ which is related to $\br'$ via, $x'=x''$, $y'=y''$, and $z'=-z''$ when the surface of the half-space is located at $z=0$. The scattering part of the Green function is then given by,
\begin{equation}
\label{eq:mult_G_scatt_slab_QS}
\begin{split}
 &\GFT_{\rm scatt,QS}\rrarg = \mp \frac{\varepsilon_2 - \varepsilon_1}{\varepsilon_2 + \varepsilon_1} \GFT_{\rm hom,QS} \left(\br,\br''\right).
\end{split}
\end{equation}


%

\end{document}